\documentclass{vldb}

\newcommand{\bbN}{\ensuremath{\mathbb{N}}}

\newcommand{\NX}{\ensuremath{\bbN[X]}}

\newcommand{\sM}{\!_M}

\newcommand{\sK}{\!_K}

\newcommand{\supp}{\mbox{supp}}

\newcommand{\eqdef}{\equiv}

\usepackage{times,amsmath,epsfig}
\usepackage{xcolor}
\usepackage{graphicx}
\usepackage{amssymb}
\usepackage{amsmath}
\usepackage{graphicx}
\usepackage{latexsym}
\usepackage{epstopdf}
\usepackage{multicol}
\usepackage{caption}
\usepackage{subcaption}
\usepackage{url}
\usepackage[vlined,ruled,commentsnumbered]{algorithm2e}
\usepackage{changepage}
\usepackage{tikz}
\usetikzlibrary{positioning,chains,fit,shapes,calc}
\usepackage{multirow}
\usepackage{array}
\newcolumntype{P}[1]{>{\centering\arraybackslash}p{#1}}
\newcolumntype{M}[1]{>{\centering\arraybackslash}m{#1}}%
\definecolor{myblue}{RGB}{80,80,160}
\definecolor{mygreen}{RGB}{80,160,80}
\DeclareTextFontCommand{\emph}{\em}
\DeclareTextFontCommand{\texttt}{\tt}
\newcommand{\exampleName}{$K$-example}
\newcommand{\trio}{Trio(X)}

\newcommand{\why}{Why(X)}

\newcommand{\BX}{B[X]}
\newcommand{\Ex}{$Ex$}
\newcommand{\ExOut}{O}
\newcommand{\ExIn}{I}

\newcommand{\systemName}{{\tt QPlain}}

\newcommand{\daniel}[1]{}
\newcommand{\amir}[1]{}

\newtheorem{theorem}{Theorem}[section]

\newtheorem{proposition}[theorem]{Proposition}

\newtheorem{example}[theorem]{Example}

\newtheorem{definition}[theorem]{Definition}

\SetCommentSty{mycommfont}

\makeatletter
\let\@copyrightspace\relax
\makeatother

\let\OLDthebibliography\thebibliography
\renewcommand\thebibliography[1]{
    \OLDthebibliography{#1}
    \setlength{\parskip}{0pt}
    \setlength{\itemsep}{0pt plus 0.3ex}
}

\begin{document}

%\setcopyright{acmcopyright}

\title{Query By Provenance}

\author{Daniel Deutch and Amir Gilad \\ Tel Aviv University}

\maketitle

\begin{abstract}

To assist non-specialists in formulating database queries, multiple
frameworks that automatically infer queries from a set of examples
have been proposed. While highly useful, a shortcoming of the
approach is that if users can only provide a small set of examples,
many inherently different queries may qualify, and only some of
these actually match the user intentions. Our main observation is
that if users further {\em explain} their examples, the set of
qualifying queries may be significantly more focused. We develop a
novel framework where users explain example tuples by choosing input
tuples that are intuitively the ``cause" for their examples. Their
explanations are automatically ``compiled" into a formal model for
explanations, based on previously developed models of {\em data
provenance}. Then, our novel algorithms infer conjunctive queries
from the examples and their explanations. We prove the computational
efficiency of the algorithms and favorable properties of inferred
queries. We have further implemented our solution in a system
prototype with an interface that assists users in formulating
explanations in an intuitive way. Our experimental results,
including a user study as well as experiments using the TPC-H benchmark, indicate the effectiveness of our solution.
\end{abstract}

%To capture explanations,
%we leverage previously developed models of {\em data provenance}, and

%and in particular their ``embedding" in the model of provenance
%semirings. An important advantage is that the obtained problem
%definition is generic and allows plugging-in explanation models of
%different levels of detail and granularity. We highlight several
%modeling and computational challenges in the context of the problem,
%and address them to develop efficient algorithms that infer
%conjunctive queries from examples and their explanations. We show
%the computational efficiency of the algorithms and favorable
%properties of inferred queries through a theoretical analysis as
%well as an experimental study using the TPC-H benchmark. Our
%experiments indicate that even when shown only few examples and
%their explanations, our system succeeds in ``reverse engineering"
%many highly complex queries.

%\end{abstract}

\section{Introduction}
%\daniel{CQ or $CQ^{neq}$?}

It has long been acknowledged that writing database queries in a
formal language is a cumbersome task for the non-specialist.
Different solutions have been proposed to assist users in this
respect; a prominent approach (see e.g. \cite{qbo,joinQueries,Shen})
allows users to provide examples of output tuples, based on which
the intended query is automatically inferred. This approach can be
highly effective if the examples provided by the user are plenty and
representative. But coming up with such a set of examples is highly
non-trivial, and unless this is the case, the system would
be unable to distinguish the true intention of the user %true intentions
from other qualifying queries.

As a simple illustration, consider a user planning to purchase
airline tickets for a trip. She has rather specific requirements:
the trip should include five countries in South America, visiting
each for a week and staying in Bolivia in the third week and in
Argentina in the fourth, in time for meetings she has scheduled.
After viewing a list of border crossings (see Table \ref{relR1}),
she concludes that Argentina and Brazil would serve as good
end-points for the trip, and so would Peru and Paraguay. Since
airfare to these particular destinations is quite expensive, she is
interested in viewing additional recommendations. However, based
only on these two examples of output tuples, there are many
inherently different queries that yield them as a subset of their
results, and there is no reasonable way to distinguish between these
queries. For instance, the trivial query copying the content of
Table \ref{relR1} also yields these two tuples.

Intuitively, if users would provide some form of ``explanations"
for their examples, it could guide the system in identifying
the actual intended query. The explanations should on the one hand
be simple enough so that non-experts are able to specify them (and
in particular their specification should be much easier than query
formulation), and on the other hand be informative enough to
allow inference of the underlying query. Continuing our running
example, an explanation for a pair of end-points involves a
description of actual trips that the user has in mind, and are
compatible with the example end-points. This would in turn limit the
queries of interest to those that not only include the example
output, but rather do so based on criteria that are compatible with
the explanation.

We propose in this paper a novel framework for learning queries
from examples {\em and explanations for these examples}, a problem that we
refer to as {\em query-by-explanation}. The backend of the framework is based on a formal model for explanations,
namely that of {\em provenance semirings} \cite{GKT-pods07}, a formal problem statement that intuitively involves ``reverse-engineering" queries from their provenance, and efficient algorithms for the problem in multiple variants. Importantly, since users can not be expected to understand complex notions of provenance, the framework includes an intuitive Graphical User Interface through which users specify explanations, by essentially dragging-and-dropping relevant input tuples (the system further assists them in this task). The provided explanations are automatically compiled to formal provenance and fed to the algorithms. The effectiveness of the solution is shown through
extensive experiments, including a user study.

Our solution comprises of the following
components.

\paragraph*{A Formal Model for Explanations (Section \ref{sec:prelim})} We first need a formal notion of
explanations to be attached to examples. In this respect, we note
that multiple lines of work have focused on the ``reverse" problem
of the one we consider here, namely that of {\em explaining query
    results}. The basic idea in all of these works is to associate with each
output tuple $t$ some description of the input tuples that ``cause" $t$ to appear in the output, i.e. they are used by the
query in a derivation that yields $t$. Different models vary in the granularity of
explanations. For instance, the {\em why-provenance} \cite{why} of
$t$ is a {\em set of sets} of input tuples, where each set includes
the tuples that have been used in a single derivation. The
provenance polynomials model (\NX\ in \cite{GKT-pods07}) essentially
extends why-provenance to account for multiplicity: each monomial of
a provenance polynomial includes the {\em annotations} (intuitively
identifiers, for our purpose of use) of tuples participating in a
single derivation. Exponents are used to capture that a tuple was
used multiple times, and coefficients capture multiple derivations
based on the same set of variables. Importantly, \cite{Greenicdt09}
has shown that these and other models may be captured through the
{\em provenance semirings} model \cite{GKT-pods07}, via different
choices of semirings. We use here the provenance semirings model as
the underlying model for explanations, and examine the effect of
different semiring choices.

\paragraph*{Query-By-Explanation (Section \ref{sec:model})}
We then formally define the novel problem of learning queries from
examples and their explanations (termed query-by-explanation).
Examples are simply output tuples, and explanations are, formally,
instances of provenance attached to them. We formally define what it
means for a query to be consistent with examples and their
explanations. Intuitively, we want a query that, when evaluated with
respect to the input database does not only yield the specified
example tuples, but also its derivation of these tuples is
consistent with the prescription made by the explanation. This is formalized by
leveraging the inclusion property \cite{Greenicdt09} of relations
annotated with elements of {\em ordered semirings}. Basing our
formal construction on these foundations allows for a ``clean",
generic, problem definition.
%
%
%Furthermore, basing our problem definition on the formal
%grounds of the provenance literature has two main benefits:
%
%\begin{itemize}
%   \item It allows to formally define what it means for a query to be consistent with an
%   example with explanations. Intuitively, we want a query that, when
%   evaluated with respect to the input database does not only yield the
%   specified example tuples, but also whose computation of these tuples
%   is consistent with the prescription made by the explanation. We
%   formalize the notion of consistency with an explanation by
%   leveraging the inclusion property of relations annotated with
%   elements of {\em ordered semirings}, introduced in
%   \cite{Greenicdt09}. The same tool allows us to also define an {\em
%       inclusion minimal} query, namely one whose computation is a ``tight
%   fit" to the given explanation.
%
%   \item Different users are able to give explanations in different
%   levels of granularity. As we will demonstrate, our generic model
%   allows such flexibility, corresponding to different choices of
%   provenance models.
%\end{itemize}

We then study the query-by-explanation problem, for Conjunctive Queries (CQs; this is a quite standard choice in this context, see e.g. \cite{Shen,joinQueries,Psallidas}) and for
different semirings used for explanations. As we discuss below (Section \ref{sec:imp}), users do not directly specify explanations in a any semiring model (and in fact do not even need to be aware of these models),
but rather only need to understand the intuitive notion of ``cause", which naturally corresponds to the reasons they had in mind when choosing examples.
Still, the user specification can be of varying levels of details, which we show to correspond to different choices of semirings.

%\daniel{next few lines here or not?}
%On the other hand, as we discuss below (Section \ref{sec:interface}), our abstraction is such that users do
%not have to understand or even be aware of the different provenance
%models. They only need to understand the intuitive notion of
%``cause", which naturally corresponds to the reasons they had in mind when choosing examples.

%Each such model requires different level of detail
%from the user, and corresponds to a different semiring, all under
%the same framework. We make a quite standard choice in this context
%(see e.g. \cite{Shen,joinQueries,Psallidas}) of further restricting
%our attention to Conjunctive Queries ($CQ$s).

\paragraph*{Learning from Detailed Explanations (Section \ref{sec:NX})} We start by assuming that a detailed form of explanation is given; formally, here we capture explanations as {\em provenance polynomials},
elements of the $\NX$ semiring. In our example it means that for
each trip the system is aware of all border crossings, including
multiple occurrence of the same crossing, but not of the order in
which they take place (order is abstracted away in
semiring provenance). Also note that not all explanations (trips, in our example) need
to be specified. Technically, a key to generating a consistent query
in this case is to ``align" provenance annotations
appearing in different monomials, eventually mapping them to
constructed query atoms. Indeed, we show that given a permutation
over all annotations appearing in each explanation (formally a
monomial in the provenance polynomial), we can efficiently construct
a corresponding atom for each location of the permutation, or
declare that none exists. However, an algorithm that exhaustively
traverses all such permutations would be prohibitively inefficient
(EXPTIME both in the monomials size and in the number of
examples). Instead, we design an efficient algorithm that is careful
to traverse combinations whose number is only exponential in the
arity of the output relation, and polynomial in the number
of examples and in the provenance size. We further adapt it to find inclusion-minimal queries.

\paragraph*{Relaxing the Level of Detail (Section \ref{sec:trio})} With $\NX$ provenance, we have assumed complete knowledge of some (but maybe not all) derivations that support the example tuple. There may still be a mismatch between the intuition of explanation that the
users have in mind and that of derivations expressed in the $\NX$
semiring. This mismatch is reflected in the existence of {\em
multiplicities} in $\NX$, i.e. multiple occurrences of the same tuple in an
explanation as well as the same explanation occurring multiple
times. Such multiplicities are usually due to the technical
operation of the query, and then they may not be specified in the
user-provided explanation. To this end, we note that the model of
{\em why-provenance} \cite{why} (captured by the $Why(X)$ semiring of \cite{Greenicdt09}) is oblivious to such multiplicities. We show that learning queries from such explanations is more
cumbersome: there may be infinitely many queries that lead to the
same $Why(X)$ provenance (with different multiplicities that are abstracted away). To this end, we prove a small world property, namely
that if a consistent query exists, then there exists such query of
size bounded by some parameters of the input. The bound by itself
does not suffice for an efficient algorithm (trying all $\NX$
expressions of sizes up to the bound would be inefficient), but we
leverage it in devising an efficient algorithm for the $Why(X)$
case. We then complete the picture by
showing that our solutions may be adapted to explanations specified in other provenance models for which a semiring-based interpretation was given in
\cite{Greenicdt09}.

%\paragraph*{Section \ref{sec:lin}} Both $\NX$ and $\trio$ provenance require the user to distinguish between joint and alternative use of tuples in a derivation. In some settings this distinction may be difficult for non-experts to make. In these cases, the explanation may take the form of lineage; namely the user only specifies a set of tuples contributing/relevant to each particular output tuple. Again, this is challenging from our perspective since it allows multiple ways in which a query may be specified while still using all tuples in joint/alternative ways. \daniel{Say something intelligent here}

\paragraph*{Implementation Details (Section \ref{sec:imp})} We have implemented our solution in a prototype
system called \systemName, that allows users to specify examples and
explanations through an intuitive GUI. Users formulate explanations
by simply dragging-and-dropping input tuples that serve as support
for their examples. Importantly, our GUI assists users in
identifying those tuples by limiting attention to ``relevant" such
tuples: first, only input tuples that have some values in common with
the provided example tuple are proposed (intuitively these will
correspond to query atoms that contribute to the head); in
subsequent steps, the system will also propose tuples that share common values with
tuples already chosen to appear in the explanation (intuitively
these will e.g. correspond to atoms used as join conditions). The
explanations are automatically compiled to expressions in the
appropriate semiring -- intuitively the least expressive of the
supported semirings whose expressive power suffices to capture the
supplied explanations --- without requiring any user awareness of
semiring model. Finally, the relevant algorithm is  invoked to
return a query of interest.

\paragraph*{Experiments (Section \ref{sec:exp})} We have conducted an extensive
experimental study of our system, to asses the feasibility of
forming explanations, the quality of inferred queries (even when very few examples and explanations are provided), and the
computational efficiency of our algorithms. To this end, we have
first conducted a user study, requesting users to provide examples
and explanations both for some complex pre-defined tasks as well as
for tasks of their choice, with respect to the IMDB movies dataset. Users were successful in forming examples and
explanations in the vast majority of the cases, and a small number of examples (up to 4) was typically sufficient for the system to infer the underlying query. Then, we have
further studied the system's ability to ``reverse-engineer" TPC-H
queries as well as highly complex join queries presented as baseline
in \cite{joinQueries}. The queries include multiple joins and
self-joins, with up to 8 atoms and up to 60 variables. In the vast majority of the cases, our algorithms converged to the
underlying query after having viewed only a small number of
examples. Last, further experiments indicate the computational efficiency of
our algorithms.

We survey related work in Section \ref{sec:related} and conclude in
Section \ref{sec:conc}.

\section{Preliminaries}
\label{sec:prelim}

%In this paper, we consider a particular class of problems, namely, computing
%conjunctive queries from exmaples of input and output tuples, and a form of
%provennace. We give a brief review of the notions used throughout the paper.

 In this section we give a brief review of the basic notions we use
 throughout the paper.

\subsection{Conjunctive Queries}

We will focus in this paper on Conjunctive Queries (see e.g. \cite{hector2002database}). Fix a database schema $\mathcal{S}$ with relation names
$\{\mathcal{R}_1,...,\mathcal{R}_n\}$ over a domain $\mathcal{C}$ of
constants. Further fix a domain $\mathcal{V}$ of variables. A {\em
conjunctive query} $q$ over $\mathcal{S}$ is an expression of the
form $T(\vec{u}^{\,}) :- \mathcal{R}_1(\vec{v_1}^{\,}), \ldots, \mathcal{R}_l(\vec{v_l}^{\,})$
where $T$ is a relation name not in $\mathcal{S}$. For all $1 \leq i
\leq n$, $\vec{v_i}^{\,}$ is a vector of the form $(x_1, \ldots,
x_k)$ where $\forall 1 \leq j \leq k. \; x_j \in \mathcal{V} \cup
\mathcal{C}$. $T(\vec{u}^{\,})$ is the query head, denoted
$head(q)$, and $\mathcal{R}_1(\vec{v_1}^{\,}), \ldots,
\mathcal{R}_l(\vec{v_l}^{\,})$ is the query body and is denoted
$body(q)$. The variables appearing in $\vec{u}^{\,}$ are called the
\textit{head variables} of $q$, and each of them must also appear in
the body. We use $CQ$ to denote the class of all conjunctive
queries, omitting details of the schema when clear from context.

%\subsection{Union of Conjunctive Queries}
%A union of conjunctive queries is an expression of the form $q = q_1 \cup q_2
%\cup \ldots \cup q_m$ where for each $i \in {1, ...,m}$, $q_i \in CQ$, and for each
%$i, j \in {1, \ldots, m}$, $head(q_i)$ and $head(q_j)$ are of the same relation.
%We say that each $q_i$ is an adjunct of the query $q$.
%The set of adjuncts of $q$ is denoted by $Adj(q)$. The class of all unions of
%conjunctive queries is denoted by $UCQ$.

We next define the notion of {\em derivations} for CQs. A derivation
$\alpha$ for a query $q \in CQ$ with respect to a database instance
$D$ is a mapping of the relational atoms of $q$ to tuples in $D$
that respects relation names and induces a mapping over arguments,
i.e. if a relational atom $R(x_1, ..., x_n)$ is mapped to a tuple
$R(a_1, ..., a_n)$ then we say that $x_i$ is mapped to $a_i$
(denoted $\alpha(x_i) = a_i$). We require that a variable $x_i$ will
not be mapped to multiple distinct values, and a constant $x_i$ will
be mapped to itself. We define $\alpha(head(q))$ as the tuple
obtained from $head(q)$ by replacing each occurrence of a variable
$x_i$ by $\alpha(x_i)$.
% and a constant $x_i$ by its value.
%\end{definition}

\begin{figure}
	\begin{center}
		\small{
			\begin{tabular}{|l|}
				\hline
				\verb"trip(x, w) :- route(x, y), route(y, Bolivia), "\\
				\verb"route(Bolivia, Argentina), route(Argentina, w)" \\
				\hline
			\end{tabular}
		} \vspace{-5mm}
	\end{center}
	\caption{$Q_{real}$} \label{fig:qreal}
	\vspace{-5mm}
\end{figure}

\begin{table}[!htb]
%\begin{minipage}[b]{1\linewidth}
\centering \small
\begin{tabular}{| c | c | c |}
\hline prov. & A & B \\
\hline a & Argentina & Brazil \\
\hline b & Brazil & Bolivia \\
\hline c & Bolivia & Argentina \\
\hline d & Peru & Colombia \\
\hline e & Colombia & Bolivia \\
\hline f & Argentina & Colombia \\
\hline g & Peru & Paraguay \\
\hline h & Argentina & Paraguay \\
\hline
\end{tabular}
%\caption*{}%\label{relR1}

%\end{minipage}%
%\begin{minipage}[b]{.5\linewidth}
%\centering
%\begin{tabular}{| c | c | c |}
%\hline prov. & A & B \\
%\hline i & Germany & Belgium \\
%\hline j & France & Belgium \\
%\hline k & Belgium & Germany \\
%\hline l & Hungary & Slovakia \\
%\hline m & Poland & Slovakia \\
%\hline n & Slovakia & Hungary \\
%\hline
%\end{tabular}
%\caption*{}%\label{relR1}
%\end{minipage}\vspace{-10mm}
%\vspace{-4mm}
\caption{Relation $route$}\label{relR1} %\vspace{-4mm}
\end{table}

\vspace{-4mm}
\begin{example}
\label{ex:prov}
Reconsider our example from the Introduction of a
user planning a trip with specific requirements. Her logic may be
captured by the query $Q_{real}$ presented in Fig. \ref{fig:qreal} ($route$ is depicted in
Table \ref{relR1}). Intuitively, the query outputs all possible endpoints of a trip that
includes visits to five countries, the third and fourth of which being Bolivia and Argentina resp.
Now, consider the result tuple $trip(Argentina,Brazil)$. It is
obtained through two different derivations: one that maps the atoms
to four distinct tuples of $route$ to the four atoms (in order of
the atoms, these are the tuples annotated by {\bf f,e,c,a} in Table
\ref{relR1}), and one that maps the tuple annotated {\bf a} to the
first and last atoms and {\bf b,c} to the second and third
respectively. Each derivation includes the border crossings of a
trip between Argentina and Brazil that satisfies the constraints.

%a user who is interested in planning a trip, and wants it to include
%five countries in South America, visiting each for a week and
%staying in Bolivia in the third week, in time for a carnival taking
%place there. Given a relation $route$ specifying pairs of bordering
%countries, she may be interested in finding possible endpoints for
%such trips (to book appropriate airplane tickets). This

%Finally, it is important that the country
%from which Alice starts the trip will allow her to rent a car and
%drive through all the countries she plans to be in.

%\daniel{Maybe better to talk about no-fees for the rental car, which
%is not necessarily symmetric (?)}.
%Further consider the query  which aims at finding two countries: The
%first and last countries of Alice's trip; Alice can then rent a car
%from the first country and return in in the last country. Such pairs
%of countries are good candidates from which Alice can begin and end
%her road trip. The query output (the $trip$ relation) along with the
%provenance for its tuples, is depicted in Table
%\ref{outRel}.% and \ref{outRel2}.

\end{example}

\subsection{Provenance}

%The tracking of {\em provenance} as explanation to query results has
%been extensively studied in multiple lines of work, with different
%models and different positions in the tradeoff between preciseness
%and provenance size. The work of \cite{Greenicdt} have shown that
%many different such models may be captured using the {\em semiring
%approach} \cite{GKT-pods07}. This unifying perspective is useful in
%our development, and we recap it next for three particular such
%models: (full) provenance polynomials, trio \cite{trio} and lineage
%\cite{lin}.

The tracking of {\em provenance} to explain query results has been
extensively studied in multiple lines of work, and
\cite{Greenicdt09} has shown that different such models may be
captured using the {\em
semiring approach} \cite{GKT-pods07}. We next overview several aspects of the approach that we will use in our formal framework.

%The idea is to capture the
%possible derivations of each tuple, while employing different
%equivalence axioms in each concrete choice of model. We next
%overview the approach and different provenance models under the
%perspective it gives.

 %The algebraic structure of commutative semiring
%\cite{AlgebraBook} was shown in \cite{GKT-pods07} to fit as a basis
%for provenance tracking. We next briefly recap it.

\paragraph*{Commutative Semirings} A {\em commutative monoid} is an algebraic structure $(M,+_{\sM},0_{\sM})$ where $+_{\sM}$ is an associative and
commutative binary operation and $0_{\sM}$ is an identity for
$+_{\sM}$. A {\em commutative semiring} is
then a structure $(K,+_{\sK} ,\cdot_{\sK},0_{\sK},1_{\sK})$
where $(K,+_{\sK} ,0_{\sK})$ and $(K,\cdot_{\sK},1_{\sK})$ are
commutative monoids, $\cdot_{\sK}$ is distributive over $+_{\sK} $,
and $a\cdot_{\sK}0_{\sK} = 0\cdot_{\sK} a = 0_{\sK}$.

\paragraph*{Annotated Databases}  
We will capture provenance through the notion of databases whose tuples are associated (``annotated") with elements of a commutative semiring.
For a schema $\mathcal{S}$ with
relation names $\{\mathcal{R}_1,...,\mathcal{R}_n\}$, denote by
$Tup(\mathcal{R}_i)$ the set of all (possibly infinitely many)
possible tuples of $\mathcal{R}_i$. 
\begin{definition} [adapted from \cite{GKT-pods07}]
A {\em $K$-relation} for a relation name $\mathcal{R}_i$ and a
commutative semiring $K$ is a function
$R:Tup(\mathcal{R}_i) \mapsto K$ such that its {\em support} defined
by $\supp(R) \eqdef \{t \mid R(t)\neq 0\}$ is finite. We say that
$R(t)$ is the annotation of $t$ in $R$. A $K$-database $D$ over a
schema $\{\mathcal{R}_1,...,\mathcal{R}_n\}$ is then a collection of
$K$-relations, over each $\mathcal{R}_i$.
\end{definition}

Intuitively a $K$-relation maps each tuple to its annotation. 
We will sometimes use $D(t)$ to denote the
annotation of $t$ in its relation in a database $D$. We furthermore say that a $K$-relation is {\em abstractly tagged}
if each tuple is annotated by a distinct element of $K$ (intuitively, its identifier).
 
\paragraph*{Provenance-Aware Query Results} We then define Conjunctive
Queries as mappings from $K$-databases to $K$-relations. Intuitively we define the annotation (provenance) of an output tuple as a combination of annotations of input tuples. This combination is based on the query derivations, via the
intuitive association of alternative derivations with the semiring
``$+$" operation, and of joint use of tuples in a derivation with
the ``$\cdot$" operation. 

\begin{definition} [adapted from \cite{GKT-pods07}]
\label{def:basicprov} Let $D$ be a $K$-database and let $Q \in CQ$,
with $T$ being the relation name in $head(Q)$. For every tuple $t
\in T$, let $\alpha_{t}$ be the set of derivations of $Q$ w.r.t. $D$
that yield $t$. $Q(D)$ is defined to be a $K$-relation $T$ s.t. for
every $t$, $T(t)=\sum_{\alpha \in \alpha_{t}}\prod_{t' \in
Im(\alpha)}D(t')$. $Im(\alpha)$ is the image of $\alpha$.
\end{definition}

Summation and multiplication in the above definition are done in an
arbitrary semiring $K$. Different semirings give different
interpretations to the operations, as we next illustrate with two important semirings.

\paragraph*{Provenance Polynomials (\NX)} The most general form of
provenance for positive relational algebra (see \cite{GKT-pods07})
is the {\em semiring of polynomials with natural numbers as coefficients}, namely $(\NX,+,\cdot,0,1)$. The idea
is that given a set of basic annotations $X$ (elements of which may
be assigned to input tuples), the output of a query is represented
by a sum of products as in Def. \ref{def:basicprov}, with only the
basic equivalence laws of commutative semirings in
place. Coefficients serve in a sense as ``shorthand" for multiple
derivations using the same tuples, and exponents as ``shorthand" for
multiple uses of a tuple in a derivation.

Many additional forms of provenance have been proposed in the
literature, varying in their level of abstraction and the details
they reveal on the derivations. We leverage here the work of
\cite{Greenicdt09} that has shown that multiple such provenance
forms may be captured through the semiring model, with the
appropriate choice of semiring. We next show an impotrant such model.

\vspace{-1mm}
\paragraph*{Why(X)} A natural approach to provenance tracking,
referred to as {\em why-}provenance \cite{why}, capturing each
derivation as a {\em set} of the annotations of tuples used in the
derivation. The overall why-provenance is thus a {\em set of such
sets}. As shown (in a slightly different way) in \cite{Greenicdt09},
this corresponds to using provenance polynomials but without ``caring" about exponents and coefficients.
Formally, consider the function $f:\NX
\mapsto \NX$ that {\em drops all coefficients and exponents} of its
input polynomial. We then introduce a congruence relation defined by
$P_1 \equiv P_2$ if $f(P_1) = f(P_2)$. $Why(X)$ is then
defined as the quotient semiring of $\NX$ under this congruence
relation (i.e. two equivalent polynomials are indistinguishable).

\begin{example}

The provenance-aware result (for \NX\ and \why) of evaluating $Q_{real}$ over the
relation $route$ is shown in Table \ref{outRel}.  Reconsider for example
the tuple $trip(Argentina, Brazil)$. Recall its two derivations
shown in Example 2.1. Consequently, its ``exact" (\NX)
provenance is $f \cdot e \cdot c \cdot a + a^2 \cdot c \cdot b$.
Each summand corresponds to a derivation, and recall that each
derivation stands for an alternative suitable trip that starts at
Argentina and ends at Brazil. Note that the provenance includes a
specification of the bag of tuples used in each derivation, in no
particular order (multiplication is commutative). If we alternatively store \why-provenance, we still have
summands standing for alternative derivations (trips), but we
further lose track of exponents, i.e. the number of times each tuple
was used (as well as multiple identical derivations, if such exist).
The why-provenance here is $f \cdot e \cdot c \cdot a + a \cdot c
\cdot b$. Note that it still specifies the border crossings made
during the trip, but we do not know that a border was crossed twice.

In general, two trips may include the same border crossings, but in
different order (e.g. (Bolivia-Argentina-Bolivia-Brazil-Bolivia) and
(Bolivia-Brazil-Bolivia-Argentina-Bolivia), if the corresponding
tuples are present in the database). In $\NX$ provenance, the corresponding monomial would have appeared with coefficient $2$; this coefficient would have been omitted in Why(X) provenance. 

%Finally, with {\em
%lineage}, we store the set of tuples that have been used in {\em
%some} derivation, which is $\{a,b,c,e,f\}$ in this case.
%Intuitively, it details the relevant input tuples (here all relevant
%border crossings), but does not distinguish between alternatively
%and jointly used tuples (here it means it does not group the border
%crossings according to the trips they appear in). \daniel{Make sure
%the example is convincing.}
\end{example}

\begin{table}[!htb]
\vspace{-1mm}
\begin{minipage}{1\linewidth}
\centering \small
\begin{tabular}{| c | c | c | c | c | c |}
\hline A & B & \NX & $Why(X)$ \\
\hline Argentina & Brazil & $f \cdot e \cdot c \cdot a+ a^2 \cdot c \cdot b$ & $ f \cdot e \cdot c \cdot a+ a \cdot c \cdot b$  \\
\hline Peru & Paraguay & $d \cdot e \cdot c \cdot h$ & $d \cdot e \cdot c \cdot h$ \\
\hline
\end{tabular}
\caption{Relation $trip$}\label{outRel}
\end{minipage}
\vspace{-6mm}
\end{table}

%So far, we have reviewed different variations of polynomials. A trio is a form of

%\paragraph*{Lin(X)}

%Throughout the paper, we will only consider the input database that
%appears in the {\exampleName}, and denote it by $D^{in}$.

%\daniel{Other candidates: PosBool[X] (boolean expressions), B[X]
%(drop coefficients), Why[X]. Need to at least mention them, would be
%good if we can say something interesting. }

%\daniel{Need to define annotated databases. DISCUSS.}

%Given a CQ query $Q$, an annotated database $D \in \mathcal{DB}$,
%and an output tuple $t \in Q(D)$, we use $prov_{Q,D}(t)$ to denote
%the provenance of $t$ with respect to $Q$ and $D$. When unclear from
%context, we will also specify the kind of provenance we refer to.

\section{Query-By-Explanation}
\label{sec:model}

We define in this section the problem of learning queries from
examples and their explanations. We first introduce the notion of
such examples, using provenance as explanations.

\begin{definition}[Examples with explanations]\label{provExample}
Given a semiring $K$, a {\em $K$-example} is a pair $(I,O)$ where $I$ is an abstractly-tagged $K$-database called the {\em input} and $O$ is a $K$-relation called the {\em output}.
\end{definition}

%Our definition assumes a single input database and multiple examples
%for desired output;  This fits our use-cases where users are
%interested in retrieving more results that resembles their examples
%with respect to a given database, rather than re-running the
%inferred query on multiple databases. A variant where multiple
%input-output pairs are given is discussed in Section
%\daniel{TODO:forward pointer}.

%We typically cannot assume that the given example is complete.
%Without explanations, i.e. in standard query-by-output setting (see
%e.g. \cite{}), this simply means that the output example may include
%only a subset of the tuples of the actual output instance. This is
%generalized to our settings as follows:

Intuitively, annotations in the input only serve as identifiers, and those in the output serve as explanations -- combinations of annotations of input tuples contributing to the output. 

We next define the notion of a query being consistent
with a $K$-example. In the context of query-by-example,
a query is consistent if its evaluation result
includes all example tuples (but maybe others as well). We resort to
\cite{Greenicdt09} for the appropriate generalization to the
provenance-aware settings:

%For a given {\exampleName}, a matching query is one whose evaluation
%result includes all output tuples (but maybe others as well), and
%whose derivation involves the specified provenance information (but
%maybe more). Both are captured by the notion of order on
%$K$-relations:
%
%
%\paragraph*{Annotated Relations Containment} We aim to use
%annotated relations as examples, but we typically cannot assume that
%the given example is complete. Without explanations, i.e. in
%standard query-by-output setting (see e.g. \cite{}), this simply
%means that the output example may include only a subset of the
%tuples of the actual output instance. Here again we resort to
%\cite{Greenicdt09} for the appropriate generalization.

\begin{definition} \cite{Greenicdt09}
Let $(K,+_{K},\cdot_{K},0,1)$ be a semiring and define $a \leq_{K}
b$ iff $\exists c.~ a+_{K} c = b$. If $\leq_{K}$ is a (partial)
order relation then we say that $K$ is naturally ordered.

Given two $K$-relations $R_1,R_2$ we say that $R_1 \subseteq_{K}
R_2$ iff \\
$\forall t. R_1(t) \leq_{K} R_2(t)$.

\end{definition}

 Note that if $R_1\subseteq_{K} R_2$ then in particular $supp(R_1) \subseteq supp(R_2)$, so the notion of containment w.r.t.
a semiring is indeed a faithful extension of 
 ``standard" relation containment. In terms of provenance, we note that for \NX and \why, the natural order corresponds to inclusion of monomials: $p_1 \leq
 p_2$ if every monomial in $p_1$ appears in $p_2$. The order relation has different interpretations in other
 semirings, still fitting the intuition of a partial explanation.

We are now ready to define the notion of consistency with respect to
a $K$-example, and introduce our problem statement. Intuitively, we
look for a query whose output is contained in the example output,
and for each example tuple, the explanations provided by the user
are ``reflected" in the computation of the tuple by the query.

\begin{definition} [Problem Statement]
%Let $p,p'$ be two provenance polynomials. We say that a provenance
%$p$ is {\em contained} in $p'$, and denote $p \leq p'$, if every
%monomial of $p$ occurs in $p'$.
\label{def:problem}

Given a {\em $K$-example} (I,O) and a conjunctive query $Q$ we say that $Q$ is consistent with respect to the example if
$O\subseteq_{K} Q(I)$. {\em K-CONSISTENT-QUERY} is the problem of finding a consistent
query for a given $K$-example.

\end{definition}

The above definition allows multiple conjunctive queries
to be consistent with a given $K$-example. This is in line with the
conventional wisdom in query-by-example; further natural desiderata w.r.t. to the query, and are studied in Section 4.3.

%Note that since all provenance models that we have introduced may be
%captured by polynomials, the definition of containment applies to
%each of them. For Lin(X) we haven't explicitly given the polynomial
%construction; containment in Lin(X) is simply set containment.
%\daniel{Not very nice. TBD: Recheck T.J.'s paper, surely there is an
%interpretation. Need to give it}
%
%\begin{example}
%
%\daniel{Removed BX. TBD: fix to current example}
%
%We demonstrate provenance containment using Example
%\ref{queryExample}. Consider the polynomial \NX $p' = b\cdot c +
%a^2$. It contains the polynomial $p = b\cdot c$ since every monomial
%in $p$ appears in $p'$ thus $p \leq p'$. Polynomial containment in
%\BX and Trio(X) is very similar to \NX.  We now exemplify polynomial
%containment in Lin(X) provenance. Let $p = \{a\}$ and $p' = \{a, b,
%c\}$ so $p \leq p'$ since every label in $p$ also appears in $p'$
%(the set $p$ is contained in the set $p'$).
%\end{example}

\begin{figure}
	\begin{center}
		\small{
			\begin{tabular}{|l|}
				\hline
				\verb"trip(x, y) :- route(x, z), route(w, y), "\\
				\verb"route(t, r), route(k, l)" \\
				\hline
			\end{tabular}
		} \vspace{-5mm}
	\end{center}
	\caption{$Q_{general}$} \label{fig:qgen}
	\vspace{-5mm}
\end{figure}

We next demonstrate the notion of consistent queries with respect to
a given {\exampleName}.

\begin{example}\label{ex:consExact}
%\daniel{TODO:replace with demo example}
Consider Table \ref{outRel}, now treated as an \NX-example. Each
monomial corresponds to a trip that fits the constraints that the
user has in mind, serving as an explanation that the user has
provided for the trip end-points. Consistent queries must derive the example tuples in the ways specified in the polynomials (and possibly in additional ways). The query $Q_{real}$ from Figure %Example \ref{ex:prov} 
\ref{fig:qreal} is of course a consistent query with respect to it,
since it generates the example tuples and the provenance of each of
them according to $Q_{real}$ is the same as that provided in the
example. $Q_{real}$ is not the only consistent query; in particular the query $Q_{general}$ presented in Fig. \ref{fig:qgen}
is also consistent (but note that this particular query is not
minimal, see further discussion in Section \ref{sec:NX}).
\end{example}

%\begin{multline*}
%$$Q_{general}:trip(x,y) :- route(x,z), route(w,y),\\
%route(t,r), route(k,l)$$
%\end{multline*}

%We claim that $Q$ from Example \ref{ex:prov} is a consistent query
%w.r.t it. To show this, we show that $Q$ holds Definition
%\ref{def:problem}. $Q$ has 4 atoms in the body as is the size of the
%monomials in the provenance, and each monomial defines an assignment
%of $Q$ that will derive the corresponding tuple in the same row as
%the monomial. For instance, by assigning $f$ to $route(x,y)$, $e$ to
%$route(y,Bolivia)$, $c$ to $route(Bolivia,z)$ and $a$ to
%$route(z,w)$, we can derive the fact $trip(Argentina,Brazil)$. In a
%similar manner, we can receive a satisfying assignment by assigning
%$a$ ($d$) to the first atom, $b$ ($e$) to the second, $c$ ($c$) to
%the third an $a$ ($h$) to the forth atom. Since for every monomial
%in the provenance, there exists an assignment that derives the
%relevant output tuple, $O\subseteq_{\mathbb{N}(X)} Q(I)$. Therefore,
%$Q$ is consistent. Note that every consistent query in the
%$\NX$-consistent is also \why-consistent, although a \why-consistent
%query is not necessarily $\NX$-consistent and we show it in Section
%\ref{sec:trio}.
%\end{example}

%is $Q: trip(x,y) :- route(x,z), route(w,y), route(t,r), route(k,l)$

In the following section we study the complexity of the above
computational problems for the different models of provenance. We
will analyze the complexity with respect to the different facets of
the input, notations for which are provided in Table
\ref{notations}.

\begin{table}[!htb]
\centering
\begin{tabular}{| c | l | c |}
\hline \Ex & {\exampleName} \\
\hline $\ExIn$ & Input database  \\
\hline $\ExOut$ & Output relation and its provenance \\
\hline $m$ & Total number of monomials \\
\hline $k$ & Number of attributes of the output relation \\
\hline $n$ & (Maximal) Number of elements in a monomial \\
\hline
\end{tabular}
\vspace{-2mm}
\caption{Notations}\label{notations}
\vspace{-3mm}
\end{table}

\section{Learning from \NX\ explanations}
\label{sec:NX}

We start our study with the case where the given provenance consists
of $\NX$ expressions. This is the most informative form of
provenance under the semiring framework. In particular, we note that
given the $\NX$ provenance, the number of query atoms (and the
relations occurring in them) are trivially identifiable. What
remains is the choice of variables to appear in the query atoms
(body and head). Still, finding a consistent query (or deciding that
there is no such choice) is non-trivial, as we next illustrate.

%\daniel{We know query length, the whole issue is to equate
%variables, both body to head (for safe yet consistent) and body to
%body (for minimal yet consistent)}
%\daniel{We start with the first issue (consistent but not minimal),
%then discuss minimality}

\subsection{First Try}

\IncMargin{1em}
\begin{algorithm}
    \SetKwFunction{goodSettingsAlgo}{ComputeQuery1}
    \SetKwInOut{Input}{input}\SetKwInOut{Output}{output}
    \LinesNumbered
    \Input{An N[X] example $Ex = (I, O)$}
    \Output{A consistent query $Q$ or an answer that none exists} \BlankLine

    Let $(t_1,M_1),...,(t_m,M_m)$ be the tuples and corresponding provenance monomials of $O$  \;
    Let $n$ be the size of each monomial \;
    $Perms \gets AllPermutations(O)$ \;
    \ForEach{$\pi \in Perms$}
    {
        \If{$\pi$ is inconsistent}{
        Continue to the next permutation\;
        }
        $Cover \gets \emptyset$\;
        \ForEach{attribute $A$ of $\ExOut$}
        {\label{loop:headAttr}
        \ForEach{$j<n$}
        {\label{loop:monom}
         Let $M^{j}_1,...,M^{j}_m$ be the tuples corresponding to the provenance atoms in the $j$'th place of each monomial in
         $\pi$ \;\label{line:collect}
         Let $\mathcal{R}$ denote the relation name of $M^{j}_1,...,M^{j}_m$\;
         \If{$\exists A' \in \mathcal{R} \; \forall 1 \leq i \leq m \;\; t_i.A = M^{j}_i.A'$}
                {\label{line:coverA}
                $Cover \gets  Cover \cup (A,j,A')$ \;
                }
            }
         }

     \If{all attributes of $\ExOut$ appear in $Cover$}
     {\label{line:allCovered}
        return $BuildQuery(Cover)$ \;\label{line:buildquery}
     }

    }
    Output ``No consistent query exists"\;

         %\ForEach{attribute $A'$ of the $M^{j}_{p}$ tuples}
%         {
%         \If{$M^{j}_i.A' \neq t_i.A$}
%         {
%         $Covered \gets false$ \;
%         }
%         }
%         \If{Covered = true}{
%             $Cover \gets  Cover \cup (A,j,A')$ \;
%         }

%        }

\caption{First (Inefficient) Try}\label{algo:FirstTry}
\end{algorithm} \DecMargin{1em}

We start by describing an inefficient algorithm (Algorithm
\ref{algo:FirstTry}) that retrieves a consistent query for a given
$\NX$ example. This will serve for explaining some main ideas of the
eventual efficient algorithm, as well as some pitfalls that need to
be avoided.

We first (line 1) ``split" the different monomials so that we obtain pairs $(t_i,M_i)$ where $t_i$ is a tuple and
$M_i$ is a monomial with coefficient $1$. To achieve that, we generate multiple copies of each tuple, one
for each monomial (a monomial with coefficient $C$ is treated as $C$ equal monomials).

Then, the goal of the algorithm is to generate query atoms while mapping the provenance
annotations to generated query atoms, in a way that is consistent
and realizes (``covers") the head attributes. To this end, a first
approach is to consider (Line 3) all possible permutations of the
annotations in every monomial (a single permutation here includes an
permutation of the annotations in $M_1$ {\em and} an permutation of
the annotations in $M_2$, and so on). Note that the need to consider multiple permutations stems from the possibility of multiple occurences of the same relation (self-joins). For each such permutation
(Lines 4-16) we try to compose a corresponding query, as follows. We
first check that the permutation is consistent (Lines 5--6) which
means that (1) for every location $j$, the atoms appearing in
location $j$ of all monomials are all annotations of tuples
appearing in the same input relation (otherwise no query atom can be
the source of all of them); and (2) every two occurrences of the
same monomial are ordered in a different way (otherwise the required
multiplicity will not be achieved). If the permutation is
consistent, we consider the head attributes one by one (Line 8), and
for each such attribute $A$ we try to find a corresponding body atom
and attribute. For that we try every location $j$ in the monomial
ordering (Line 9), and for each such location we ``collect" the
input tuples corresponding to the $j$'th atoms of all monomials
(Line 10). The head variable for $A$ may fit any attribute of the
$j$'th atom, so we need to consider every such attribute $A'$ of the
relation $R$ of the corresponding tuples (Lines 11-13; such a
relation exists due to the consistency of the permutation). This
attribute is a good fit for the head variable if this is the case
for {\em every} example monomial. If such a good fit was found, then the
variable assigned to the head attribute $A$ will appear as the $A'$
attribute of the $j$'th atom, and we continue. If all head
attributes are covered in this fashion, then we generate the
corresponding query (Lines 14-15) assigning a query atom to each
location in the ordering and placing each head variable for
attribute $A$ in the location of the covering attribute $A'$. In
contrast, if after considering all orderings, no such cover is
found, then we conclude that no consistent query exists.

\begin{table}[!htb]
\label{table:exampleperm}
\begin{minipage}{.5\linewidth}
\centering \small
\begin{tabular}{| c | c | c | c | c | c |}
\hline 1 & 2 & 3 & 4 \\
\hline f & e & c & a \\
\hline a & a & c & b \\
\hline d & e & c & h \\
\hline
\end{tabular}
\caption*{First perm.} \label{order1}

\end{minipage}%
\begin{minipage}{.5\linewidth}
\centering \small
\begin{tabular}{| c | c | c | c | c | c |}
\hline 1 & 2 & 3 & 4 \\
\hline f & a  & c & e \\
\hline a & a & c & b \\
\hline d & h & c & e \\
\hline
\end{tabular}
\caption*{Second perm.} \label{order2}
\end{minipage}
%\vspace{5mm}
\caption{Two Permutations in Ex.
\ref{ex:naturalNaive}}\label{table:exampleperm}
\vspace{-5mm}
\end{table}

\begin{example}\label{ex:naturalNaive}
Consider the three monomials in our running example (two of them
belong to the provenance of the same tuple, and are ``split" by the
algorithm). Two of the permutations are depicted in Table
\ref{table:exampleperm}. The first fails in ``covering" all head
attributes: the second output attribute consists of the values $[Brazil,
Brazil,\\Paraguay]$ (in order of the output tuples), but no index
$1\leq j\leq 4$ in the permutation is such that the input tuples
whose annotations appear in the $j$'th column of the permutation
have these values appearing in any attribute $A'$ (so the condition
in line \ref{line:coverA} is not met). In contrast, the second
permutation yields a cover for both head attributes: the first
attribute is covered for $j=1$ (via the first input attribute) and
the second attribute is now covered for $j=2$ (via the second input
attribute). Therefore, the condition in line \ref{line:allCovered}
will hold, and the algorithm will generate the query $Q_{general}$
shown in Figure \ref{fig:qgen}.%Example \ref{ex:consExact}.

\end{example}
\vspace{-4mm}
\paragraph*{Pitfalls} There presented algorithm has two pitfalls. The first is that it is prohibitively inefficient: it traverses all $n!^{m}$ possible permutations of monomials. The second pitfall is that the query
generated by the algorithm is a very general one, i.e. it does not
account for possible joins or selection criteria that may appear in
the query. In fact, as exemplified above, the query may include
``redundant" atoms, while an alternative consistent query may be
minimal.

We next address these pitfalls. We start by presenting an efficient variant of the algorithm. Then, we show
how further constraints may be inferred, to obtain a ``tight" fit to
the examples.

\subsection{An Improved Algorithm}\label{subsec:improved}
We present an alternative algorithm that avoids the exponential
dependency on $n$ and $m$. An important observation is that we can
focus on finding atoms that ``cover" the attributes of the output
relation, and the number of required such atoms is at most $k$ (the
arity of the output relation). We may need further atoms so that the
query realizes all provenance tokens (eventually, these atoms will
also be useful in imposing e.g. join constraints), and this is where
care is needed to avoid an exponential blow-up with respect to the
provenance size. To this end, we observe that we may generate a
``most general" part of the query simply by generating atoms with
fresh variables, and without considering all permutations of parts
that do not contribute to the head. This will suffice to guarantee a consistent
query, but may still lead to a generation of a too general
query; this issue will be addressed in Section 4.3.

%\IncMargin{1em}
%\begin{algorithm}
%    \SetKwFunction{goodSettingsAlgo}{SetTwoRows}
%    \SetKwInOut{Input}{input}\SetKwInOut{Output}{output}
%    \LinesNumbered
%    \Input{PEI \Ex}
%    \Output{$PM(\mathcal{E})$} \BlankLine
%    $M \gets \emptyset$\; \label{initPM}
%    Build $G(\mathcal{E}) = (V,E)$\; \label{buildGraph}
%    \ForEach {$E' \subseteq E \text{ s.t. } |E'| \leq k$}
%    {\label{EdgeSets}
%        \If {$\cup_{e \in E'} s(e) = \{1, \ldots, k\}$}
%        {\label{unionLabels}
%            \If {$\forall e_i \neq e_j \in E' \; \; e_i \cap e_j = \emptyset$}
%            {\label{isDisjoint}
%                \If {$G^\ast$ has a matching}
%                {\label{isMatching}
%                    $M.insert(E')$\;\label{insertEdges}
%                }
%            }
%        }
%    }
%
%    \Return $M$\; \label{endGoodSettingsAlgo}
%
%    \caption{SetTwoRows}\label{goodSettingsAlgo}
%\end{algorithm}
%\DecMargin{1em}

\IncMargin{1em}
\begin{algorithm}
    \SetKwFunction{goodSettingsAlgo}{ComputeQuery2}
    \SetKwInOut{Input}{input}\SetKwInOut{Output}{output}
    \LinesNumbered
    \Input{An N[X] example $Ex=(I,O)$}
    \Output{A consistent query $q$ or an answer that none exists} \BlankLine

    Let $(t_1,M_1),...,(t_m,M_m)$ be the tuples and corresponding provenance monomials of $O$  \;
    $(V,E) \gets BuildLabeledGraph((t_1,M_1),(t_2,M_2))$ \; \label{buildGraph}
    \ForEach {consistent {\em matchings} $E' \subseteq E \text{ s.t. } |E'| \leq k$}
    {\label{EdgeSets}
        \If {$\cup_{e \in E'} label(e) = \{1, \ldots, k\}$}
        {\label{line:cover}
        $Q \gets BuildQueryFromMatching(E',Ex)$ \;\label{line:buildquery2}
        \ForEach {$1<j<m$}{
         \If{not $consistent(Q,t_j,M_j)$}
         {\label{line:notconsistent}
             Go to next matching \;
         }
        }
        return $Q$ \;
    }
    }
    Output ``No consistent query exists"\;

\caption{FindConsistentQuery (N[X])}
\label{algo:Efficient}
\end{algorithm} \DecMargin{1em}

We next detail the construction, shown in Algorithm
\ref{algo:Efficient}. Again, we separate monomials similarly to
Algorithm 1, but this time not duplicating monomials with
coefficient greater than $1$ (see a dedicated treatment of
coefficients below). We then start by picking two tuples and
monomials (see below a heuristic for making such a pick) and denote
the tuples by $t_1$ and $t_2$ and their provenance by $M_1=a_1 \cdot
... \cdot a_n$ and $M_2=b_1 \cdot ... \cdot b_n$ respectively. Our
goal is to find all ``matches" of {\em parts} of the monomials so
that all output attributes are covered. To this end, we define (Line
2) a full bipartite graph $G= (V_1 \cup V_2, E)$ where each of $V_1$
and $V_2$ is a set of $n$ nodes labeled by $a_1, \ldots, a_n$ and
$b_1, \ldots, b_n$ respectively. We also define labels on each edge,
with the label of $(a_i,b_j)$ being the set of all attributes that
are {\em covered} by $a_i,b_j$, in the same sense as in lines 12-13
of Algorithm \ref{algo:FirstTry}: an output attribute $A$ is covered
if there is an input attribute $A'$ whose value in the tuple
corresponding to $a_i$ ($b_j$), matches the value of the attribute
$A$ in $t_1$ (respectively $t_2$).

We then (lines 3-4) find all matchings, {\em of size $k$ or less},
that cover all output attributes; namely, that the union of sets
appearing as labels of the matching's edges equals $\{1,...,k\}$. As
part of the matching, we also specify which subset of the edge label
attributes is chosen to be covered by each edge (the number of such
options is exponential in $k$). It is easy to observe that if such a
cover (of any size) exists, then there exists a cover of size $k$ or
less. We further require that the matching is consistent in the
sense that the permutation that it implies is consistent.

For each such matching we generate (line 5) a ``most general query"
$Q$ corresponding to it, as follows. We first consider the matched
pairs $a_i,b_j$ one by one, and generate a query atom for each pair.
This is done in the same manner as in the last step of Algorithm
\ref{algo:FirstTry}, except that the query generation is done here
based on only $k$ pairs of provenance atoms, rather than all $n$
atoms. To this end, we further generate for each provenance token
$a_i$ that was not included in the matching a new query atom with
the relation name of the tuple corresponding to $a_i$, and fresh
variables. Intuitively, we impose minimal additional constraints,
while covering all head attributes and achieving the required query
size of $n$.

Each such query $Q$ is considered as a ``candidate", and its
consistency needs  to be verified with respect to the other tuples
of the example (Line \ref{line:notconsistent}).  One way of doing so
is simply by evaluating the query with respect to the input, checking that the output tuples are generated, and their provenance
includes those appearing in the example. As a more efficient
solution, we test for consistency of $Q$ with respect to each
example tuple by first assigning the output tuple values to the head
variables, as well as to the occurrences of these variables in the
body of $Q$ (by our construction, they can occur in at most $k$
query atoms). For query atoms corresponding to provenance annotations that
have not participated in the cover, we only need to check that for
each relation name, there is the same number of query atoms and of
provenance annotations relating to it. A subtlety here is in handling
coefficients; for the part of provenance that has participated in the
cover, we can count the number of assignments. This number is
multiplied by the number of ways to order the other atoms (which is
a simple combinatorial computation), and the result should exceed
the provided coefficient.

%\daniel{heuristic for choosing the tuples}
\paragraph*{Choosing the two tuples} For correctness and worst case complexity guarantees, any choice of
tuples as a starting point for the algorithm (line 1) would suffice.
Naturally, this choice still affects the practical performance, and
we aim at minimizing the number of candidate matchings. A simple but
effective heuristic is to choose two tuples and monomials for which
the number of distinct values (both in the output tuple and in input
tuples participating in the derivations) is maximal.

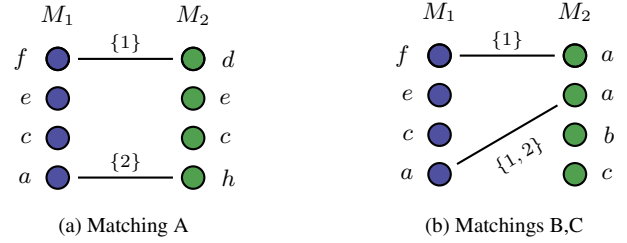
\begin{figure}[!htb]
 \centering
         \begin{subfigure}[b]{0.4\columnwidth}
                 \centering
\begin{tikzpicture}[thick,
  every node/.style={draw,circle},
  fsnode/.style={fill=myblue,node distance=.2cm},
  ssnode/.style={fill=mygreen,node distance=.2cm},
  every fit/.style={ellipse,draw,inner sep=-2pt,text width=1.25cm},
  -,shorten >= 3pt,shorten <= 3pt
]

% the vertices of U
\begin{scope}[start chain=going below,node distance=7mm]
  \node[fsnode,on chain] (1) [label=left: $f$] {};
  \node[fsnode,on chain] (2) [label=left: $e$] {};
  \node[fsnode,on chain] (3) [label=left: $c$] {};
  \node[fsnode,on chain] (4) [label=left: $a$] {};
\end{scope}

% the vertices of V
\begin{scope}[xshift=1.8cm,start chain=going below,node distance=7mm]%yshift=-0.5cm,
  \node[ssnode,on chain] (5) [label=right: $d$] {};
  \node[ssnode,on chain] (6) [label=right: $e$] {};
  \node[ssnode,on chain] (7) [label=right: $c$] {};
  \node[ssnode,on chain] (8) [label=right: $h$] {};
\end{scope}

% the set U
\node [label=above:$M_1$] {};%myblue,fit=(1) (2) (3) (4),
% the set V
\node [xshift=1.8cm,label=above:$M_2$] {};%mygreen,fit=(5) (6) (7) (8),

% the edges
\draw (1) -- (5)
node[draw=none,fill=none,font=\scriptsize,midway,above,yshift=-.2cm]
{$\{1\}$};

\draw(4)-- (8)
node[draw=none,fill=none,font=\scriptsize,midway,above,yshift=-.2cm]
{$\{2\}$};
\end{tikzpicture}
\caption{Matching A} \label{matching}
         \end{subfigure}\hfill%
         \begin{subfigure}[b]{0.4\columnwidth}
                 \centering
\begin{tikzpicture}[thick,
  every node/.style={draw,circle},
  fsnode/.style={fill=myblue,node distance=.2cm},
  ssnode/.style={fill=mygreen,node distance=.2cm},
  every fit/.style={ellipse,draw,inner sep=-2pt,text width=1.25cm},
  -,shorten >= 3pt,shorten <= 3pt
]

% the vertices of U
\begin{scope}[start chain=going below,node distance=7mm]
  \node[fsnode,on chain] (1) [label=left: $f$] {};
  \node[fsnode,on chain] (2) [label=left: $e$] {};
  \node[fsnode,on chain] (3) [label=left: $c$] {};
  \node[fsnode,on chain] (4) [label=left: $a$] {};
\end{scope}

% the vertices of V
\begin{scope}[xshift=1.8cm,start chain=going below,node distance=7mm]%yshift=-0.5cm,
  \node[ssnode,on chain] (5) [label=right: $a$] {};
  \node[ssnode,on chain] (6) [label=right: $a$] {};
  \node[ssnode,on chain] (7) [label=right: $b$] {};
  \node[ssnode,on chain] (8) [label=right: $c$] {};
\end{scope}

% the set U
\node [label=above:$M_1$] {};%myblue,fit=(1) (2) (3) (4),
% the set V
\node [xshift=1.8cm,label=above:$M_2$] {};%mygreen,fit=(5) (6) (7) (8),

% the edges
\draw (1) -- (5)
node[draw=none,fill=none,font=\scriptsize,midway,above,yshift=-.2cm]
{$\{1\}$};

\draw(4)-- (6)
node[sloped,anchor=south,auto=false,draw=none,fill=none,font=\scriptsize,midway,below,yshift=.2cm]
{$\{1, 2\}$};
\end{tikzpicture}
\caption{Matchings B,C} \label{matching2}
         \end{subfigure}
%\vspace{5mm}
\caption{Matchings for Example \ref{ex:naturalAlgo}} \vspace{-4mm}
\label{matchings}
\end{figure}

\begin{example}\label{ex:naturalAlgo}
%\daniel{TBD: the same example, now with the more efficient
%algorithm}

%\daniel{Add the other graph and matching as well, adapt the example
%text accordingly.}

Reconsider our running example. Assume that the monomials $f \cdot e
\cdot c \cdot a$ and $d \cdot e \cdot c \cdot h$ were picked for
graph generation. Figure \ref{matching} depicts a matching of size
$2$ where the first attribute of input tuples $f$ and $d$ covers the
first output attribute, and the second attribute of input tuples $a$
and $h$ covers the second output attribute. Generating a query based on this matching results in
the query $Q_{general}$ from Figure \ref{fig:qgen}. The %Example \ref{ex:consExact}
algorithm now verifies the consistency of $Q_{general}$ with respect to the third
monomial by assigning the output tuple to the head, i.e. assigning
$x$ to $Argentina$ and $y$ to $Brazil$, and returns $Q_{general}$. Two different matchings (one including only the edge (a,a) and another one including (a,a) and (f,a)) are presented in Figure
\ref{matching2}.

%If the two monomials of the first output tuple are chosen for the
%graph, it would cause a slightly different action sequence. The two
%possible matchings are $E' = \{(a,a)\}, E' = \{(a,a),(f,a)\}$
%(Figure \ref{matching2}). Consider the first matching. This matching
%will cause the generation of the query (Line
%\ref{line:buildquery2}):
%
%\begin{multline*}
%$$trip(x,y) :- route(x,y), route(w,z), route(t,r), route(k,l)$$
%\end{multline*}
%
%This query will be inconsistent with the second row of the example,
%and therefore, it will be disqualified. The second matching however,
%will again result in the generation of $Q'$.

%\daniel{merge that into the example}

%Consider the case of choosing the two monomial of the first output
%tuple in our running example. There will be two possible matchings:
%$(f,a), (a,a)$ and $(a,a)$. However, by choosing the monomial from
%the second row as one of the two monomials from which the graph is
%built, we can immediately disqualify a matching consisting of only
%one edge \amir{need more details?}. A generalization of this
%approach is to choose the most different two output tuples, as
%described above, and build the graph from them. In Section
%\ref{sec:exp} we show that this approach is very successful
%experimentally.\amir{remove ref. to section \ref{sec:exp}?}

\end{example}

%$trip(x,y) :- route(x,z), route(w,y),\\route(t,r), route(k,l)$ again
%resulting in the query $Q'$.

\vspace{-4mm}
\paragraph*{Complexity} The algorithm's complexity is $O(n^{2k} \cdot
m)$: at most $n^{k}$
matchings are considered; for each matching, a single query is
generated, and consistency is validated in
$O(n^{k})$ for each of the $m$ example tuples.
We have avoided the exponential factor in $n$ and $m$ and instead the exponential factor only involves $k$, which is much
smaller in practice. Can we do even better? We can show that if $P \neq NP$, there is no algorithm running in time polynomial in $k$.
\begin{proposition}
Deciding the existence of a consistent query with respect to a given
$\NX$ example is NP-hard in $k$.
\end{proposition}

\subsection{Achieving a tight fit}
\label{subsec:minimize}

We have now developed an efficient algorithm for deciding the
existence of a consistent query, and computing one if exists. Still,
as exemplified above, a downside of the algorithm is that the
generated query is very general. A natural desideratum (employed in
the context of ``query-by-example"), is that the query is
``inclusion-minimal". This notion extends naturally to
$K$-databases.

\begin{definition}
A consistent query $Q$ (w.r.t. a given $K$-example $Ex$) is \emph{inclusion-minimal} if for every query $Q'$ such that $Q'
\subsetneq_{K} Q$ (i.e. for every $K$-database $D$ it holds that
$Q'(D) \subseteq_{K} Q(D)$, but not vice-versa), $Q'$ is not consistent w.r.t. $Ex$.
\end{definition}

To find inclusion-minimal queries, we next refine
Algorithm~\ref{algo:Efficient} as follows. We do not halt when
finding a single consistent query, but instead find all of those
queries obtained for some matching. For each consistent query $Q$,
we examine queries obtained from $Q$ by (i) equating variables and
(ii) replacing variables by constants where possible (i.e. via an
exact containment mapping \cite{AHV}). We refer to both as
\emph{variable equating}. To explore the possible combinations of
variable equatings, we use an algorithm inspired by data mining
techniques (e.g.,~\cite{Gunopulos1997data}): in each iteration, the
algorithm starts from a minimal set of variable equatings that was
not yet determined to be (in)consistent with the example. E.g., in
the first iteration it starts by equating a particular pair of
variables. The algorithm then tries, one-by-one, to add variable
equatings to the set, while applying transitive closure to ensure
the set reflects an equivalence relation. If an additional equating
leads to an inconsistent query, it is discarded. Each equatings set
obtained in this manner corresponds to a homomorphism $h$ over the
variables of the query $Q$, and we use $h(Q)$ to denote the query
resulting from replacing each variable $x$ by $h(x)$.

%To find inclusion-minimal queries, we next refine Algorithm 2 as
%follows. We do not halt when finding a single consistent query, but
%instead find all of those queries obtained for some matching. For
%each consistent query $Q$, we will examine queries obtained from $Q$
%by equating variables and replacing variables by constants where
%possible (i.e. via an exact containment mapping \cite{AHV-1995}). A
%difficulty is that naively going through all such options would be
%inefficient. Instead, we consider the lattice defined (but not
%explicitly generated) as follows: its elements are all sets of pairs
%of variables occurring in $q$, as well as pairs of the form $(x,c)$
%where $x$ is a variable and $c$ is a constant (it is enough to
%consider constants co-occurring, in the position of $x$ in the first
%two examples). The partial order of the lattice is set inclusion. If
%$S \subseteq S'$ we will say that $S'$ is an ``extension" of $S$.
%Each node defines an homomorphism $h$ over the variables of $q$,
%that groups all variables occurring together in a pair and maps
%variables to constants when they are paired together (we will
%generate the lattice on-the-fly, generating only nodes whose set of
%pairs defines an equivalence relation, so that we can compute the
%transitive closure of the equalities). We use $h(q)$ to denote the
%query obtained from $q$ through replacing every variable $x$ by
%$h(x)$.

Importantly, by equating variables or replacing variables by
constants we only impose further constraints and obtain no new
derivations. In particular, the following result holds, as a simple
consequence of Theorem~7.11 in \cite{Greenicdt09} (note that we must
keep the number of atoms intact to be consistent with the
provenance):

\begin{proposition}
Let $Q$ be a CQ over a set of variables
$\mathcal{V}$. Let $h:\mathcal{V} \mapsto \mathcal{V} \cup
\mathcal{C}$ be a homomorphism. For every $\NX$-example $Ex$, if $Q$
is not consistent with $Ex$, then neither is $h(Q)$.
\end{proposition}

Consequently, the algorithm finds a \emph{maximal set of variable
equatings} that is consistent with the query, by attempting to add
at most $O(k'^2)$ different equatings, where $k'$ is the number of
unique attributes in the body of $Q$. We record every query that was
found to be (in)consistent -- in particular, every subset of a
consistent set of equatings is also consistent -- and use it in the
following iterations (which again find maximal sets of equatings).

\paragraph*{Checking for consistency}
This check may be done very efficiently for query atoms that
contribute to the head, since we only need to check that equality
holds for the provenance annotations assigned to them. For other
atoms we no longer have their consistency as a given and
in the worst case we would need to examine all matchings of these query atoms
atoms to provenance annotations.

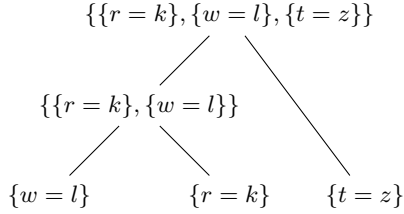
\begin{figure}
\centering
\begin{tikzpicture}[scale=0.6]
  \node (max) at (0,4) {$\{\{r = k\}, \{w = l\}, \{t = z\}\}$};
  \node (a) at (-2,2) {$\{\{r = k\}, \{w = l\}\}$};
  %\node (c) at (2,2) {$\{\{r = k\}, \{t = z\}\}$};
  \node (d) at (-4,0) {$\{w = l\}$};
  \node (e) at (0,0) {$\{r = k\}$};
  \node (f) at (3,0) {$\{t = z\}$};
  \draw (d) -- (a) -- (max)
  (f) -- (max);
  \draw[preaction={draw=white, -,line width=6pt}] (a) -- (e);
\end{tikzpicture}
\caption{Part of the lattice in Example \ref{ex:compareVars}}
\label{lattice}
\end{figure}

\begin{example}\label{ex:compareVars}
Reconsider our running example query $Q_{general}$; a part of the
lattice is depicted in Figure \ref{lattice}. The algorithm starts by
considering individually each pair of variables as well as pairs of
variables and constants co-appearing in the two output tuples or in
the tuples used in their provenance. In our example, when
considering the lattice element $\{r=k\}$, the algorithm will find
that the query $Q_{r=k}$ (i.e. $Q_{general}$ after equating $r$ and
$k$), is still consistent. Next, the algorithm will find that
equating $l,w$ in $Q_{r=k}$ also yields a consistent query so it
will proceed with $Q_{r=k,w=l}$, etc. Of course, multiple steps may
yield the same equivalence classes in which case we perform the
computation only once. Three more equalities, namely $t=z$,
$r=`Bolivia'$ and $w=`Argentina'$ may be employed, leading to the
``real" query $Q_{real}$ in Fig. \ref{fig:qreal}. Any further step
with respect to $Q_{real}$ leads to an inconsistent query, and so it
is returned as output.

\end{example}
\vspace{-3mm}
\paragraph*{Choosing a Single Query to Output} For each consistent
query found by Algorithm 2, there may be multiple inclusion-minimal
queries obtained in such a manner (though the number of such queries
observed in practice was not very large, see Section 7). If we wish
to provide a single query as output, we may impose further criteria.
A particularly important consideration here is the ``syntactic"
minimality (in the sense of Chandra and Merlin \cite{ChandraMerlin})
of the inferred query. This is a desirable feature of the inferred
query, but there is a subtlety in this respect when considering
provenance: $\NX$ provenance is not preserved under syntactic
minimization (in particular, we may get less atoms than specified in
the provenance). We can thus check candidate queries for syntactic
minimality, and prefer those that satisfy it (if any). Testing for
minimality via the algorithm of \cite{ChandraMerlin} is quite costly
(exponential time in the query size), but we run it only on
consistent inclusion-minimal queries whose number is small. Finally,
if multiple inclusion-minimal and syntactically-minimal consistent
queries are obtained, a natural and simple heuristic that we employ
is to prefer a query with the least number of unique variables.

\section{Learning from \why-Examples}
\label{sec:trio}

%\daniel{Need to talk about ``why" also}

We next study the problem of learning queries from $\why$-examples.
Such explanation is often easier for users to provide, but is in
turn more challenging for query inference.

\subsection{Challenges and First try}

A natural approach is to reduce the problem of learning from a
$\why$-example to that of learning from an $\NX$-example. Recall
that the differences are the lack of coefficients and the lack of
exponents. The former is trivial to address (we simply do not need
to check that coefficients are realized), but the latter means
that we do not know the number of query atoms. Surprisingly, attempting to bound the query size by the size of the
largest monomial fails:

\begin{proposition}
\label{prop:naivefails} There exists a $\why$ example for which
there is no consistent conjunctive query with $n$ atoms
(recall that $n$ is the length of the largest monomial),
but there exists a consistent conjunctive query with more atoms.
\end{proposition}

It is however not difficult to show a ``small world" property, based
on a looser bound.

\begin{proposition}
\label{prop:triobound} For any $\why$ example, if there exists a
consistent query then there exists a consistent query with $k+
r\cdot(n-1)$ atoms or less, where $r$ is the number of distinct
relation names occurring in the provenance monomials.
\end{proposition}

Intuitively, there are at most $k$ atoms contributing to the head.
The worst case is when only one ``duplicated" annotation contributes
to the head, and then for each example there are at most $n-1$
remaining annotations. If the query includes a single relation name
($r=1$), then a query with at most $n-1$ more atoms would be
consistent. Otherwise, as many atoms may be needed for each relation
name. Together with our algorithm for $\NX$, Proposition
\ref{prop:triobound} dictates a simple algorithm that exhaustively
goes through all $\NX$ expressions that are compatible with the
$\why$ expressions appearing in the example, and whose sizes are up
to $n+k$. This, however, would be highly inefficient. We next
present a much more efficient algorithm.

\subsection{An Efficient Algorithm}

An efficient algorithm for finding CQs consistent
with a given $\why$ provenance is given in Algorithm
\ref{algo:trioEfficient}. The idea is to traverse the examples one
by one, trying to ``expand'' (by adding atoms) candidate queries
computed thus far to be consistent with the current example. We
start (Line 1), as in the $\NX$ case, by ``splitting'' monomials if
needed so that each tuple is associated with a single monomial. We
maintain a map $\mathcal{Q}$ whose values are candidate queries, and
keys are the parts of the query that contribute to the head, in a
canonical form (e.g. atoms are lexicographically ordered). This will
allow us to maintain only a single representative for each such
``contributing part'', where the representative is consistent with
all the examples observed so far. For the first step (line
\ref{line:init}) we initialize $\mathcal{Q}$ so that it includes
only $(t_1,M_1)$ (just for the first iteration, we store an example
rather than a query). We then traverse the remaining examples one by
one (line \ref{line:loopRows}). In each iteration $i$, we consider
all queries in $\mathcal{Q}$; for each such query $Q$, we build a
bipartite graph (line \ref{line:buildGraph}) whose one side is the
annotations appearing in $M_i$, and the other side is the {\em atoms
of $Q$}. The label on each edge is the set of head attributes
covered jointly by the two sides: in the first iteration this is
exactly as in the $\NX$ algorithm, and in subsequent iterations we
keep track of covered attributes by each query atom. Then, instead
of looking for {\em matchings} in the bipartite graph, we find (line
\ref{line:allSubgraphs}) all {\em sub-graphs} whose edges cover all
head attributes (again specifying a choice of attributes subset for
each edge). Intuitively, having $e$ edges adjacent to the same
provenance annotation corresponds to the same annotation
appearing with exponent $e$, so we ``duplicate'' it $e$ times (Lines
\ref{line:expandMonom}-\ref{line:duplicateTuple}). On the other
hand, if multiple edges are adjacent to a single query atom, we also
need to ``split'' (Lines
\ref{line:expandQuery}-\ref{line:duplicateQuery}) each such atom,
i.e. to replace it by multiple atoms (as many as the number of edges
connected to it). Intuitively each copy will contribute to the
generation of a single annotation in the monomial. Now  (line \ref{line:buildQuery}), we construct a query $Q'$ based on the
matching and the previous query ``version'' $Q$: the head is
built as in Algorithm 2, and if there were $x$ atoms not
contributing to the head with relation name $R$ in $Q$, then the
number of such atoms in $Q'$ is the maximum of $x$ and the
number of annotations in $M_i$ of tuples in $R$ that were not
matched. Now, we ``combine'' $Q'$ with $Q''$ which is the currently
stored version of a query with the same contributing atoms (lines
\ref{line:getQuery}- \ref{line:putQuery}). Combining means setting
number of atoms for each relation name not contributing to the head
to be the maximum of this number in $Q'$ and $Q''$.

\paragraph*{Complexity} The number of keys in
$\mathcal{Q}$ is exponential only in $k$; the loops thus iterate at
most $m \cdot n^{k} \cdot n^{k} \cdot (n+n^2)$ times, so the overall complexity is $O(n^{O(k)}
\cdot m)$.

%The overall complexity is $O(n^{O(k)} \cdot m)$. To observe that
%this holds, note that the number of ``splits" is bounded to at most
%$k-1$ otherwise we reach query size of $n+k$
%or more. %By starting with a query of length $n$ (i.e. of the same
%length as the largest monomial), since each atom has $k-1$ options
%for splitting (not splitting, split once, split twice and so on),
%the total number of possibilities for this query to change is
%$n^{O(k)}$.
%
%The first graph can have $O(n^{2k})$ covering subgraphs, each
%matching a different query. Each such query has $n^{O(k)}$ possible
%splits. Thus, the total number of queries examined during the
%algorithm's run is $n^{O(k)})$. Since, each query can be checked
%with every line, the complexity of the algorithm is $O(n^{O(k)}
%\cdot m)$.

\paragraph*{Achieving a tight fit} Algorithm
\ref{algo:trioEfficient} produces a set of candidate queries, which may neither be syntactically minimal nor
inclusion-minimal. To discard atoms that are ``clearly" redundant,
we first try removing atoms not contributing to the head, and test
for consistency. We then perform the process of
inclusion-minimization as in Section \ref{subsec:minimize} (note
that $Why(X)$-inclusion was shown in \cite{Greenicdt09} to be
characterized by onto mappings which is a weaker requirement).

\IncMargin{1em}
\begin{algorithm}
	\SetKwFunction{goodSettingsAlgo}{ComputeQuery2}
	\SetKwInOut{Input}{input}\SetKwInOut{Output}{output}
	\LinesNumbered
	\Input{A $\why$ example Ex}
	\Output{A set of consistent queries (possibly empty, if none exists)} \BlankLine

	Let $(t_1,M_1),...,(t_m,M_m)$ be the tuples and corresponding provenance monomials of $Ex$  \;
	$\mathcal{Q} \gets \{NULL:(t_1,M_1)\}$ \; \label{line:init}
	\ForEach{$2 \leq i\leq m$} {\label{line:loopRows}
		\ForEach{$Q \in \mathrm{values}(\mathcal{Q})$} {\label{line:loopQueries}
			$\mathcal{Q} \gets \mathcal{Q} - \{Q\}$ \;
			$(V_1 \cup V_2, E) \gets BuildGraph(Q, (t_i,M_i))$ \;\label{line:buildGraph}
			\ForEach {{\em sub-graph} $E' \subseteq E \text{ s.t. } |E'| \leq k \text{~~and~~} \cup_{e \in E'} label(e) = \{1, \ldots, k\}$} {\label{line:allSubgraphs}
				\ForEach{provenance annotation $a$ in $M_i$} {\label{line:expandMonom}
					\If{$a$ is an endpoint of more than one edge in $E'$} {
						$E' \gets split(E',a)$ \;\label{line:duplicateTuple}
					}
				}
				\ForEach{atom $C \in Q$} {\label{line:expandQuery}
					\If{$C$ is an endpoint of more than one edge in $E'$}{
						$E' \gets split(E',C)$ \;\label{line:duplicateQuery}
					}
				}
				
				$Q' \gets BuildQuery(E',Q)$\;\label{line:buildQuery}
				$Q'' \gets \mathcal{Q}.get(contribs(Q'))$ \; \label{line:getQuery}
				$\mathcal{Q}.put(contribs(Q'),combine(Q',Q''))$
				\; \label{line:putQuery}
			}
		}
	}
	
	return $values(\mathcal{Q})$ \;\label{line:returnquery}

	\caption{FindConsistentQuery (\why)}
	
	\label{algo:trioEfficient}
\end{algorithm} \DecMargin{1em}  

\begin{example}\label{ex:algoTrioRunning}
Reconsider our running example, but now with the $\why$ provenance
given in Table \ref{outRel}. If we start from the two monomials of
the tuple $(Argentina,Brazil)$ then we generate a bipartite graph
with $V_1 = \{f,e,c,a\}$ and  $V_2 = \{a,b,c\}$, and obtain three
options for covering: $E'_1$ where the edge $(a,a)$ covers
attributes $\{1,2\}$, $E'_2$ where additionally $(f,a)$ covers
$\{1\}$, and $E'_3$ where $(a,a)$ covers $\{2\}$ and $(f,a)$ covers
$\{1\}$ (the latter two are options relevant to the same sub-graph).
When we continue with $E'_{1}$, no duplication is performed, and we
get a query $Q$ with a single $R(x,y)$ atom contributing to the
head, and three most general atoms. Then, we match $Q$ to $d\cdot
e\cdot c\cdot h$, resulting in a sub-graph matching $R(x,y)$ to both
$d,h$. This will lead to a split of the atom to $R(x,z)$ and
$R(w,y)$ that will appear in the final query, together with the
three most general atoms.  After variables equating and
removing redundant tuples, we obtain $Q_{real}$. The same query will
be obtained in different ways if choosing $E'_2$ or $E'_3$.
\end{example}

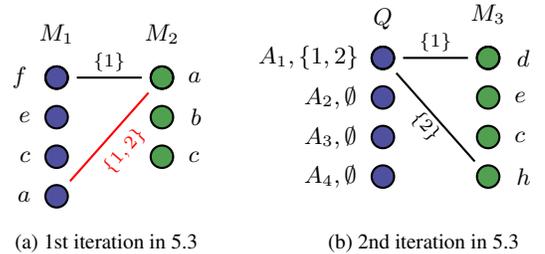
\begin{figure}[!htb]
\vspace{-3mm}
 \centering
\begin{subfigure}[b]{0.5\columnwidth}

\centering
\begin{tikzpicture}[thick,
  every node/.style={draw,circle},
  fsnode/.style={fill=myblue,node distance=.2cm},
  ssnode/.style={fill=mygreen,node distance=.2cm},
  every fit/.style={ellipse,draw,inner sep=-2pt,text width=1.25cm},
  -,shorten >= 3pt,shorten <= 3pt
]

% the vertices of U
\begin{scope}[start chain=going below,node distance=7mm]
  \node[fsnode,on chain] (1) [label=left: $f$] {};
  \node[fsnode,on chain] (2) [label=left: $e$] {};
  \node[fsnode,on chain] (3) [label=left: $c$] {};
  \node[fsnode,on chain] (4) [label=left: $a$] {};
\end{scope}

% the vertices of V
\begin{scope}[xshift=1.4cm,start chain=going below,node distance=7mm]%yshift=-0.5cm,
  \node[ssnode,on chain] (5) [label=right: $a$] {};
  \node[ssnode,on chain] (6) [label=right: $b$] {};
  \node[ssnode,on chain] (7) [label=right: $c$] {};
\end{scope}

% the set U
\node [label=above:$M_1$] {};%[myblue,fit=(1) (2) (3) (4),label=above:$M_1$]
% the set V
\node [xshift=1.4cm,label=above:$M_2$]{};%[mygreen,fit=(5) (6) (7) (8),label=above:$M_2$]

% the edges
\draw(1) -- (5)
node[draw=none,fill=none,font=\scriptsize,midway,above,yshift=-.2cm]
{$\{1\}$};

\draw[red](4)-- (5)
node[sloped,anchor=south,auto=false,draw=none,fill=none,font=\scriptsize,midway,below,yshift=.25cm]
{$\{1,2\}$};
\end{tikzpicture}
\caption{1st iteration in \ref{ex:algoTrioRunning}}
\label{subgraph1}
\end{subfigure}%
%\hfill
\begin{subfigure}[b]{0.5\columnwidth}
\hspace*{-0.8cm} \centering
\begin{tikzpicture}[thick,
  every node/.style={draw,circle},
  fsnode/.style={fill=myblue,node distance=.2cm},
  ssnode/.style={fill=mygreen,node distance=.2cm},
   every fit/.style={ellipse,draw,inner sep=-2pt,text width=1.25cm},
  -,shorten >= 3pt,shorten <= 3pt
]

% the vertices of U
\begin{scope}[start chain=going below,node distance=7mm]
  \node[fsnode,on chain] (1) [label=left: {$A_1, \{1, 2\}$}] {};%r(x,z)
  \node[fsnode,on chain] (2) [label=left: {$A_2, \emptyset$}] {};%r(w,y)
  \node[fsnode,on chain] (3) [label=left: {$A_3, \emptyset$}] {};%r(t,r)
  \node[fsnode,on chain] (4) [label=left: {$A_4, \emptyset$}] {};%r(k,l)
\end{scope}

% the vertices of V
\begin{scope}[xshift=1.4cm,start chain=going below,node distance=7mm]%yshift=-0.5cm,
  \node[ssnode,on chain] (5) [label=right: $d$] {};
  \node[ssnode,on chain] (6) [label=right: $e$] {};
  \node[ssnode,on chain] (7) [label=right: $c$] {};
  \node[ssnode,on chain] (8) [label=right: $h$] {};
\end{scope}

% the set U
\node [label=above:$Q$] {};%myblue,fit=(1) (2) (3) (4),
% the set V
\node [xshift=1.4cm,label=above:$M_3$] {};%mygreen,fit=(5) (6) (7) (8),

% the edges
\draw (1) -- (5)
node[draw=none,fill=none,font=\scriptsize,midway,above,yshift=-.2cm]
{$\{1\}$};

\draw(1) -- (8)
node[sloped,anchor=south,auto=false,draw=none,fill=none,font=\scriptsize,midway,below,yshift=.23cm]
{$\{2\}$};
\end{tikzpicture}
\caption{2nd iteration in \ref{ex:algoTrioRunning}}
\label{subgraph2}
\end{subfigure}%%
%\hfill
%\begin{subfigure}[b]{0.25\columnwidth}
%\centering
%\begin{tikzpicture}[thick,
%  every node/.style={draw,circle},
%  fsnode/.style={fill=myblue,node distance=.2cm},
%  ssnode/.style={fill=mygreen,node distance=.2cm},
%  every fit/.style={ellipse,draw,inner,sep=-2pt,text width=1.25cm},
%  -,shorten >= 3pt,shorten <= 3pt
%]
%
%% the vertices of U
%\begin{scope}[start chain=going below,node distance=7mm]
%  \node[fsnode,on chain] (1) [label=left: $a$] {};
%  \node[fsnode,on chain] (2) [label=left: $b$] {};
%\end{scope}
%
%% the vertices of V
%\begin{scope}[xshift=1.4cm,start chain=going below,node distance=7mm]
%  \node[ssnode,on chain] (3) [label=right: $c$] {};
%  \node[ssnode,on chain] (4) [label=right: $d$] {};
%\end{scope}
%
%% the set U
%\node [label=above:$Q$] {};%myblue,fit=(1) (2),
%% the set V
%\node [xshift=1.4cm,label=above:$M_2$] {};%mygreen,fit=(3) (4),
%
%% the edges
%\draw (1) to node
%[draw=none,fill=none,font=\scriptsize,midway,above,yshift=-.2cm]
%{$\{1\}$} (3);
%
%\draw (1) to node [draw=none,fill=none,font=\scriptsize,near
%end,above,yshift=-.24cm] {$\{2\}$} (4);
%
%\draw (2) to node [draw=none,fill=none,font=\scriptsize,near
%start,above,yshift=-.23cm] {$\{3\}$} (4);
%
%\end{tikzpicture}
%\caption{\ref{ex:trioAlgo} subgraph} \label{subgraph3}
%\end{subfigure}
%%\vspace{5mm}
\caption{Subgraphs for Example \ref{ex:algoTrioRunning}} \vspace{-4mm}
\label{matchings}
\end{figure}

\vspace{-1mm}
\subsection{Additional Semirings}
To complete the picture, we next show how to adapt our algorithms to explanations taken from semirings (presented in \cite{Greenicdt09}) additional to $\NX$ and $\why$.

%The algorithmic solutions, however, depend on the choice of semiring. We have provided algorithms for $\NX$, $\trio$ and $\Why$; to complete the picture, we breifly consider in this section additional structures that are commonly used in the context of provenance tracking, explaining how our solutions may be adapted.

\paragraph*{$\trio$ and $\BX$} Recall that $\NX$ keeps both coefficients and exponents, and $Why(X)$ drops both. Other alternatives include the $Trio(X)$ semiring where coefficients are
kept but exponents are not, and the $B[X]$ semiring where exponents
are kept but coefficients are not. For $Trio(X)$ we can employ the same algorithm
designed for $Why(X)$ (Algorithm 3), with a simple modification:
upon checking consistency of a candidate query with a tuple, we further check that there are as many
derivations that use the tuples of the monomial as dictated by the
coefficient (as done in Section \ref{subsec:improved}). For $B[X]$
we adapt the algorithm of $\NX$, omitting the treatment of
coefficients.
\vspace{-1mm}
\paragraph*{$PosBool[X]$} The semiring of positive boolean
expressions is defined with $+$ and $\cdot$ defined as disjunction
and conjunction respectively. If the expressions are given in DNF,
then our algorithm for $Why(X)$ may be used here as well: the only
difference is the possible absorption of monomials ($a+a \cdot b
\equiv a$), but we already assume that only a subset of the
monomials are given. If the expressions are given in an arbitrary
form there is an additional (exponential time) pre-processing step
of transforming them into DNF.

Last, designing an effective solution for the lineage model \cite{lin} is left as an important task for future work. 

\vspace{-2mm}
\section{Implementation}\label{sec:imp}

\begin{figure}[]
	\begin{center}
		\includegraphics[trim = 0mm 0mm 0mm 0mm, clip = true, width=3.6in]{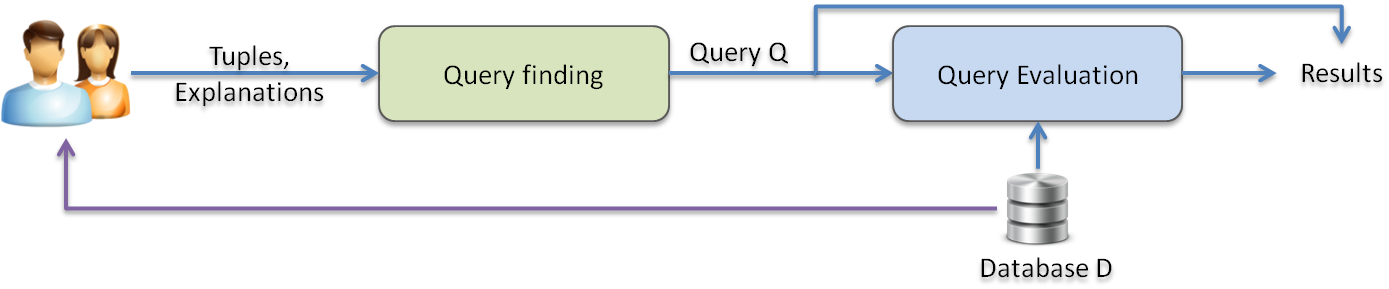}%0mm 128mm 0mm 0mmsysArch_cropped.pdf
		\vspace{-3mm} \caption{System architecture} \vspace{-1mm}
		\label{systemArchitecture}
	\end{center}
\end{figure}

\begin{figure}[]
	\centering
	\includegraphics[scale=0.3, trim={0 0 0 1.4cm}]{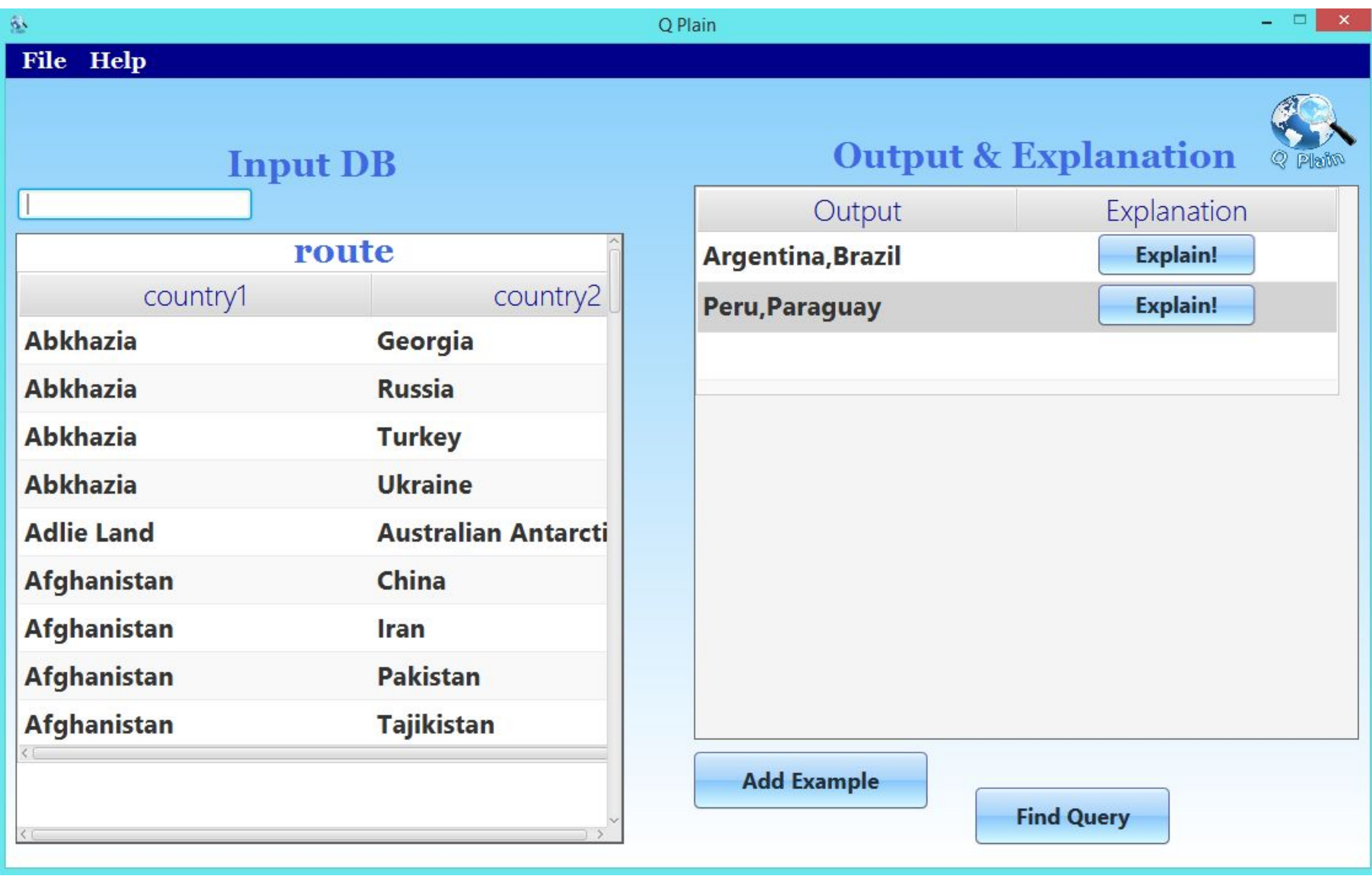}%width=3.5in
	\caption{Examples screen}
	\vspace{-4mm}
	\label{input}
\end{figure}

\begin{figure}[]
	\centering
	\includegraphics[height=1.8in]{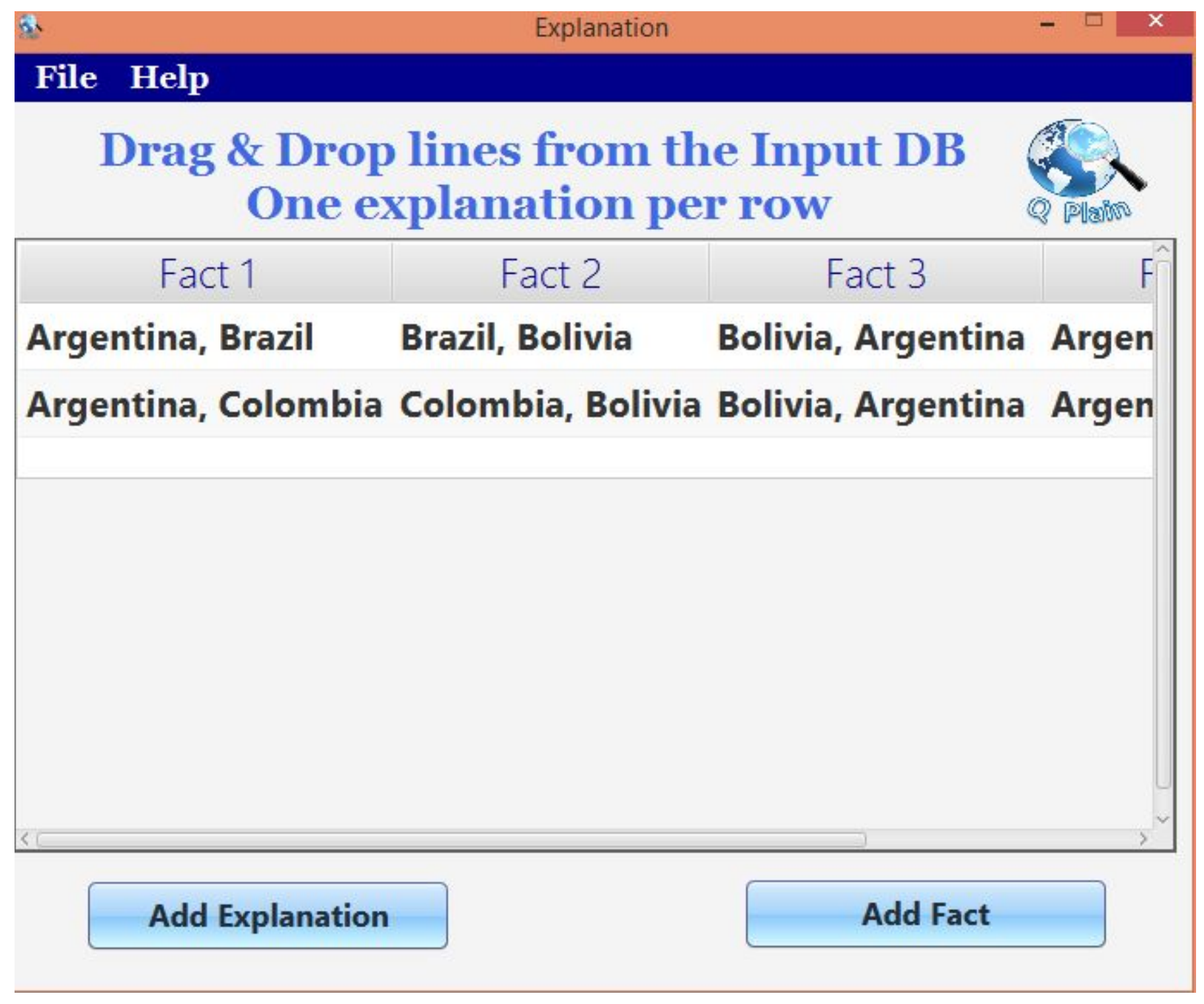}%scale=0.3
	\caption{Explanation screen}
	\vspace{-2mm}
	\label{exp}
\end{figure}

We have implemented our algorithms in an end-to-end system prototype called \systemName\ (to be demonstrated in the upcoming ICDE \cite{icde16}), implemented in JAVA with JAVAFX GUI and MS SQL server as its
underlying database management system. 
The system architecture is depicted in Figure \ref{systemArchitecture}: users load and view an input database, and then provide examples of output tuples (see Fig. \ref{input}). The users further provide explanations through a dedicated interface (see Fig. \ref{exp}). To form each explanation, users simply drag-and-drop tuples from the input database, that they intuitively view as the cause for the example tuple. Internally, the annotations of tuples chosen for each explanation are combined to form a monomial in the provenance expression.

Importantly, the system assists users in choosing the tuples to form an explanations. To this end, we observe that (unless the intended query is degenerate in that it has only constants in the head) each monomial (explanation) must contain at least one annotation of a tuple that shares at least one value with the example tuple. Consequently, we first only ask the user to choose an explanation tuple out of the input tuples that share at least one value with her example tuple. For instance, in our running example, the user is first asked to choose either the first or last crossing point of the trip between the end-points she has given (see more examples in Section \ref{sec:userstudy}). Once a first explanation tuple is given, the proposals for the following one include again tuples that share values with the example tuple, but now also tuples that share values with the given explanation tuple (this corresponds to a join condition, e.g. a second crossing point in our example), and so on.

Once the user explanations are in place, the system ``compiles" it into a provenance expression of the corresponding semring. The choice of semiring is done automatically; the system assumes the ``simplest" semiring that can accommodate the user-provided explanations: if there are no repetitions then $Why(X)$ is assumed; if there are repetitions of explanations but not of tuples within an explanation then $Trio(X)$ is assumed, etc. 

Finally, the tuples and explanations (in their ``compiled" form) are directed to either of the algorithms presented in the previous sections (according to the identified semiring). The algorithm outputs a query and this query is evaluated with respect to the underlying database, showing the user the full output set, and allowing her to add examples and re-run the inference process as needed.

\section{Experiments}
\label{sec:exp}

%\begin{figure}[]
%\begin{center}
%\includegraphics[trim = 0mm 0mm 0mm 0mm, clip = true, width=3.7in]{sysArch_cropped.pdf}%0mm 128mm 0mm 0mm
%\vspace{-2mm} \caption{System architecture} \vspace{-3mm}
%\label{systemArchitecture}
%\end{center}
%\end{figure}

%\amir{reason for different percentages in the same query: slight
%variation in the input DB in order to create the example}

We have performed an experimental study whose goals were to assess: (1) can users provide meaningful explanations for their examples? (2) once (a small number of) examples and explanations are in place, how effective is the system in inferring queries? (3) how efficient are the algorithms in terms of execution time?

To this end, we have performed two kinds of experiments. The first is a user study based on the IMDB movies database (a part of its schema is depicted in Fig. \ref{scehma}); the second is based on the actual output and provenance of the benchmark queries in \cite{joinQueries}. All experiments were performed on Windows 8, 64-bit, with 8GB of RAM
and Intel Core Duo i7 2.59 GHz processor. We next describe both experiments and their results.

\subsection{User Study}
\label{sec:userstudy}
\begin{figure}[]
	\begin{center}
		\includegraphics[trim = 0mm 0mm 0mm 0mm, clip = true, width=3in]{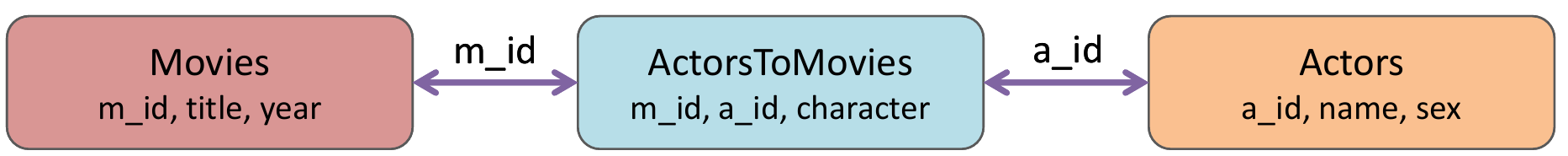}
		\vspace{-2mm} \caption{Partial IMDB schema} \vspace{-3mm}
		\label{scehma}
	\end{center}
\end{figure}

We have examined the usefulness of the system to non-expert users. To this end, we have loaded the IMDB database to \systemName\ (see a partial schema in Figure \ref{scehma}) and have presented each of the tasks in Table \ref{tasks} to 15 users. We have also allowed them to freely choose tasks of the likings, resulting in a total of 120 performed tasks. The intended queries are presented in Table \ref{queries}, where the relation {\em atm} stands for ActorsToMovies.        

% The experiment then consisted of two steps: first, we have asked users to perform the tasks in Table \ref{tasks} by providing the system with examples and explanations; then we have allowed users to choose questions of their liking, and again interact with the system to have these questions answered. The study consisted of 15 participants \amir{cur. have 13}, where each was asked to perform 7 tasks, combining to a total of 105 system interactions. In addition, we allowed the users to come up with their own tasks. 

%
%In this study, we aimed to understand the feasibility of a non-expert user providing output examples and explanations for the system.
%For the user study, we have used \systemName\ as described in Section \ref{sec:imp} with the IMDB dataset \cite{imdb}. The user tasks are shown in Table \ref{tasks}.  

%\paragraph*{Study Settings}

%To give an example, the user would copy the output value/values from the table (e.g., copy the actor name from the actor table) and drag the tuples he think were needed to explain why he chose this value (e.g., the tuple that states that the actor played in Pulp Fiction). All intended queries consist of five atoms or less. 
%similar to the sample given in Figure \ref{fig:imdb}
%Once the user has finished with the examples, he would press on the ``Find Query'' button which would allow the system to find the query using the aforementioned algorithms and evaluate it using IRIS \cite{iris}.

\begin{table}[!htb]
	\centering \scriptsize
	\begin{tabular}{l | p{7.7cm} }
		\hline & {\bf Task} \\ 
		\hline
		1. & Find all actresses \\
		2. & Find all movies that came out in 1994 \\
		3. & Find all of Al Pacino's movies \\
		4. & Find the entire cast of Pulp Fiction \\
		5. & Find all documentary films (by having character ``Himself'') \\
		%6. & Find pairs of actors who played in two movies together \\
		6. & Find the actors who played in all the 3 movies of The Lord Of The Rings \\
		7. & Choose an actor and two movies he played in, and find all actors that played with him in these two movies \\
		%
		%Instead of 4: 9. Choose an actor and two movies he played in, and
		%find all actors that played with him in these two movies.
		%
		%Sometimes give the example and sometimes not. Maybe analyze what
		%would have happended without the explanation.
		%
		%Did they succeed, and how difficult was it?
		
		\hline
	\end{tabular}
	\caption{User Tasks}\label{tasks}
	\vspace{-2mm}
\end{table}

%Order of questions in Iris Experiment. Do not delete!
%\begin{table}[!htb]
%    \centering
%    \begin{tabular}{l  l }
%        \hline & {\bf Task} \\
%        \hline
%        1. & Find all actresses \\
%        2. & Find all of Al Pacino's movies \\
%        3. & Find the entire cast of Pulp Fiction \\
%        %4. & Find the actors who played both in Pulp Fiction and in Reservoir Dogs \\
%        5. & Find pairs of actors who played in two movies together \\
%        6. & Find all movies that came out in 1994 \\
%        7. & Find the actors who played in all the 3 movies of Jurassic Park \\
%        8. & Find all documentary films (by character) \\
%        9. & Choose an actor and two movies he played in, and find all actors that played with him in these two movies \\
%%
%%Instead of 4: 9. Choose an actor and two movies he played in, and
%%find all actors that played with him in these two movies.
%%
%%Sometimes give the example and sometimes not. Maybe analyze what
%%would have happended without the explanation.
%%
%%Did they succeed, and how difficult was it?
%
%        \hline
%    \end{tabular}
%    \caption{User Tasks}\label{tasks}
%\end{table}

\begin{table}[!htb]
	\centering \scriptsize
	\begin{tabular}{l |  p{7.7cm} }
		\hline & {\bf Queries} \\ 
		\hline 
		1. & ans(t) :- actors(mid, t, 'F') \\
		2. & ans(t) :- movies(mid, t, '1994') \\ %actorsToMovies
		3. & ans(t) :- atm(mid, aid, c), actors(aid, 'Pacino, Al', 'M'), movies(mid, t, y) \\
		4. & ans(n) :- atm(mid, aid, c), movies(mid, 'Pulp Fiction', '1994'), actors(aid, n, s) \\
		5. & ans(t) :- atm(mid, aid, 'Himself'), actors(aid, n, 'M'), movies(mid, t, y) \\
		6. & ans(n) :- atm(mid1, aid, c), atm(mid2, aid, c), atm(mid3, aid, c), actors(aid, n, s), movies(mid1, 'The Fellowship of the Ring', '2001'), movies(mid2, 'The Two Towers', '2002'), movies(mid3, 'The Return of the King', '2003') \\
		\hline
	\end{tabular}
	\caption{Queries (No. 7 depends on user choices, omitted)}\label{queries}
	\vspace{-4mm}
\end{table}

%Overall the vast majority of users (80\%) were successful in properly formulating examples and explanations for all tasks.

In 111 out of the 120 cases, including all cases of freely chosen tasks, users were successful in specifying examples and explanations, and the interface of \systemName\ was highly helpful in that.

\begin{example}
	%\daniel{We need an example that illustrates all parts: how the system's proposals helped users in formulating explanations, what were the examples and the explanations, how was the query inferred and what would happen if not. Build upon the following. }
	Consider task 6 in Table \ref{tasks}. The examples are actor names; for every chosen actor name, \systemName\ has proposed as explanations only the tuples that included this actor name as one of its values.  In particular, the tuples of  $ActorsToMovies$ corresponding to movies in which the actor has played were proposed. Knowing the underlying task, it was natural for users to choose the relevant movies, namely the three ``Lord Of The Rings" movies. Once they did that, the tuples of these movies (and not of any other movie) in the $Movies$ relation has appeared as proposals, allowing their easy selection. A similar case is that of task 3: once a movie is specified as an example, the system proposes its actors as possible explanations. The choice of Al Pacino as an explanation reveals the underlying intention.  
\end{example}

In turn, explanations were crucial for allowing the system to focus on the intended query, even with very few examples (in all cases users provided at most 4 examples). 

\begin{example}
	
	%\amir{add more examples from mail and explain in detail}
	
	Re-consider task 6, and now assume that explanations are unavailable. There would be no way of distinguishing the underlying query from, for instance, a much simpler one that looks for the actors of a {\em single} Lord of the rings. More generally, join conditions (or even the fact that a join took place) are in many cases only apparent through the explanations: another example is task 3, where examples are simply movie names that typically have many other characteristics in common in addition to having Al Pacino acting in them. In addition to revealing join conditions, explanations help to learn the query constants: for instance, in task 4, without explanations, the resulting query could have returned all actors who played in a movie with Quentin Tarantino.
	
	%unless this explanation was provided, the system could infer a different query, e.g., a query that returns all movies who have male actors in them. 
\end{example}

Out of the 9 cases where users failed to provide examples and explanations, 5 involved difficulties in providing any example matching the task, and 4 involved errors in providing explanations. Out of the remaining 111 cases, in 98 cases \systemName\ has inferred the actual underlying query, and in the remaining 13 cases, it has inferred a ``more specific" query (i.e. with extra constants). This for instance happened when all actors given as examples were males. We next further analyze the accuracy of \systemName.

\subsection{Accuracy}

To further examine the effectiveness of the 
approach, we have used the system to ``reverse 
engineer" multiple complex queries. The queries are Q1--Q6 from \cite{joinQueries} as well as (modified, to drop aggregation and arithmetics) the TPC-H queries TQ2--TQ5, TQ8 and TQ10. The queries have 2--8 atoms, 18--60 variables, and multiple instances of self-joins (We show Q6 for illustration in Figure \ref{fig:query}; the reader is referred to \cite{joinQueries,tpc} for the other queries).  
We have evaluated each query using a proprietary provenance-aware
query engine \cite{icde15}, and have then sampled random fragments (of a given size that we vary) of the output database and its provenance (we have tried both \NX\ and Why(X)), feeding it to our system. In each experiment we have gradually added random
examples until our algorithm has retrieved the original query. This
was repeated 3 times. We report (1) the {\em worst-case} (as
observed in the 3 executions) number of examples needed until the
original query is inferred, and (2) for fewer examples (i.e. before
convergence to the actual query), the differences between the
inferred queries and the actual one (we report the differences
observed in the ``worst-case" run of the experiment).

\begin{table*}[!htb]
\centering\small
\begin{tabular}{| M{1.5cm} | M{3cm} | p{12cm} |}
\hline \textbf{Query} & \textbf{Worst-case number of examples to learn the original query} & \textbf{Difference between original and inferred queries for fewer examples}\\
\hline Q1 (TQ3) & 14 & \noindent\parbox[c]{\hsize}{Inferred Query includes an extra join on a ``status" attribute of two relations. Only 2--3 values are possible for this attribute, and equality often holds.}\\%or 'P' for orderstatus {\em orders} and {\em lineitem} atoms in attributes orderstatus and linestatus since that the\\ two attr. get values that are either 'F', 'O'  many times \daniel{Rephrase?}
\hline Q2 & 2 & \noindent\parbox[c]{\hsize}{}\\
\hline Q3 & 5 & \noindent\parbox[c]{\hsize}{For 2 examples, the inferred query contained an extra constant.
For 3 and 4 examples, it included an extra join.}\\% \daniel{actually we never discuss adding a constant! Need to.}
\hline Q4 & 19 & \noindent\parbox[c]{\hsize}{For 2 examples, the inferred query included an extra constant. For 3--18, it included an extra join on a highly skewed ``status" attribute.}\\
\hline Q5 & 11 & \noindent\parbox[c]{\hsize}{The inferred query included an extra join on a ``name" attribute.} \\
\hline Q6 & 3 &  \noindent\parbox[c]{\hsize}{The inferred query included an extra constant.}\\
\hline TQ4 & 234 & \noindent\parbox[c]{\hsize}{The inferred query included an extra join on ``orderstatus" and ``linestatus" attributes of two relations (they have two possible values). One of the original join conditions has lead to
occurrence of the same value in these attributes in the vast majority of joined tuples.} \\%Extra join between 99%'orders' and 'lineitem' atoms in attributes 'orderstatus' and 'linestatus' since both attributes get similar values in almost all derivations\\
%\hline TQ3 & 14 & same as Q1 \\
\hline TQ10 & 4 & \noindent\parbox[c]{\hsize}{The inferred query contained an extra constant.} \\
\hline TQ2 & 3 &  \noindent\parbox[c]{\hsize}{The inferred query contained an extra constant.}\\
\hline TQ5 & 3 &  \noindent\parbox[c]{\hsize}{The inferred query contained an extra constant.}\\
\hline TQ8 & 18 & \noindent\parbox[c]{\hsize}{For 2 examples, the inferred query  contained an extra constant. For 4-17 examples, the query had an extra join between a ``status" attribute of two relations.} \\
\hline
\end{tabular}
\caption{Results for the TPC-H query set and the queries from \protect\cite{joinQueries} with \NX\ provenance}\label{resultsNx}%\cite{joinQueries}
\end{table*}

\begin{table*}[!htb]
\centering\small
\begin{tabular}{| M{1.5cm} | M{3cm} | p{12cm} |}%{| c | c | l | l | l | l |}
\hline \textbf{Query} & \textbf{Worst-case number of examples to learn the original query} & \textbf{Difference between original and inferred queries for fewer examples}\\
%\hline Q1 (TQ3) & 14 & Extra join between the {\em orders} atom's orderstatus attr., to the {\em lineitem} atom's linestatus attr. due to the fact that the two attr. get similar values in most of the tuples: either 'F' or 'O' \\
\hline Q2 & 2 & \\
\hline Q3 & 5 & The inferred query for 2--4 examples did not include self-joins. \\%\begin{tabular}{@{}l@{}}\end{tabular}
\hline Q4 & 19 & \noindent\parbox[c]{\hsize}{For 2--3 examples, the inferred query did not include self-joins. For 4--18 examples, the query had an extra join on a ``status" attribute.}\\%\begin{tabular}{@{}l@{}} \end{tabular}
\hline Q5 & 13 & The inferred query for 2--12 examples did not include self-joins. \\%\begin{tabular}{@{}l@{}} \end{tabular}
\hline Q6 & 3 &  The inferred query included an extra constant.\\
%\hline TQ4 & 300 &  \begin{tabular}{@{}c@{}}Extra join between {\em orders} and {\em lineitem} atoms in attributes orderstatus \\and linestatus since that the two attr. get values that are either 'F', 'O' or 'P'\end{tabular}\\
%\hline TQ3 & 14 & for two examples, found a query with a constant that match all the DB. For all nubmer of examples up to 14, there was an extra join between the 'orders' relation's third attribute: 'orderstatus' to the lineitem relation 'linestatus' due to the fact that the two attributes get similar values in most of the tuples: either 'F' or 'O' \\
%\hline TQ10 & 4 & The query generated contained a constant\\
%\hline TQ2 & 2 &  \\
%\hline TQ5 & 2 &  \\
\hline TQ8 & 18 & \noindent\parbox[c]{\hsize}{For 2--3 examples, the inferred query contained an extra constant. For 4-17 the query had an extra join between a ``status" attribute of two relations.} \\%{\em nation} and {\em region} atoms \begin{tabular}{@{}l@{}}\end{tabular}
\hline
\end{tabular}
\caption{Results for the TPC-H query set and the queries from \protect\cite{joinQueries} containing self-joins with $Why(X)$ provenance}% 
\label{resultsTrio}
\end{table*}

The results are reported in Table \ref{resultsNx}. Observe that for some queries
the convergence is immediate, and achieved when viewing only 2--5
examples. For other queries, more examples are needed, but with one
exception (TQ4), we converge to the original query after viewing at
most 19 tuples for the different queries. For TQ4 only a very small
fraction of the output tuples reveal that an extra join should not
have appeared, and so we need one of these tuples to appear in the
sample. Furthermore, even for smaller sets of examples, the inferred
query was not ``far" from the actual query. The most commonly
observed difference involved extra constants occurring in the
inferred query (this has typically happened for a small number of
examples, where a constant has co-occurred by chance). Another type
of error was an extra join in the inferred query; this happened
often when two relations involved in the query had a binary or
trinary attribute (such as the ``status" attribute occurring in
multiple variants in TPC-H relations), which is furthermore skewed
(for instance, when other join conditions almost always imply
equality of the relevant attributes). We have also measured the precision and recall of the
output of the inferred query w.r.t. that of the original one. Obviously, when the
original query was obtained, the precision and recall were 100\%. Even when presented with fewer examples, in almost all
cases already with 5 examples, the precision was 100\% and
the recall was above 90\%. The only exception was Q5 with 75\% recall
for 5 examples.

\begin{figure}
	\begin{center}
		\scriptsize{
			\begin{tabular}{|l|}
				\hline
				\verb"ans(a, b) :- supplier(c, a, add, k, p, d, c1), "\\
				\verb"partsupp(h, c, v, j, c2), part(h, i, z, q, t, s, e, rp2, c3), "\\ 
				\verb"partsupp(h, o, x, n, c4), supplier(o, b, y, w, p2, d2, c5), "\\
				\verb"nation(k, na1, r, c6), region(r, u, c7), nation(w, na2, r, c8)"\\
				\hline
			\end{tabular}
		} \vspace{-5mm}
	\end{center}
	\caption{$Q6$} \label{fig:query}
	\vspace{-5mm}
\end{figure}

The results for $\why$ are
shown in Table \ref{resultsTrio}. For queries with no self-join, the observed results were naturally the same as
in the $\NX$ case; we thus report the results only for queries that include
self-joins (some of the queries included multiple self-joins). When presented with a very small number of examples, our algorithm was not always able to detect the
self-joins (see comments in Table 7); but the overall number of
examples required for convergence has only marginally increased with
respect to the $\NX$ case.

\begin{figure}
 \hspace*{-0.8cm}
 \centering
         \begin{subfigure}[b]{0.26\textwidth}
                 \centering
                 \includegraphics[trim=0cm 2cm 0cm 0cm, width=2.0in]{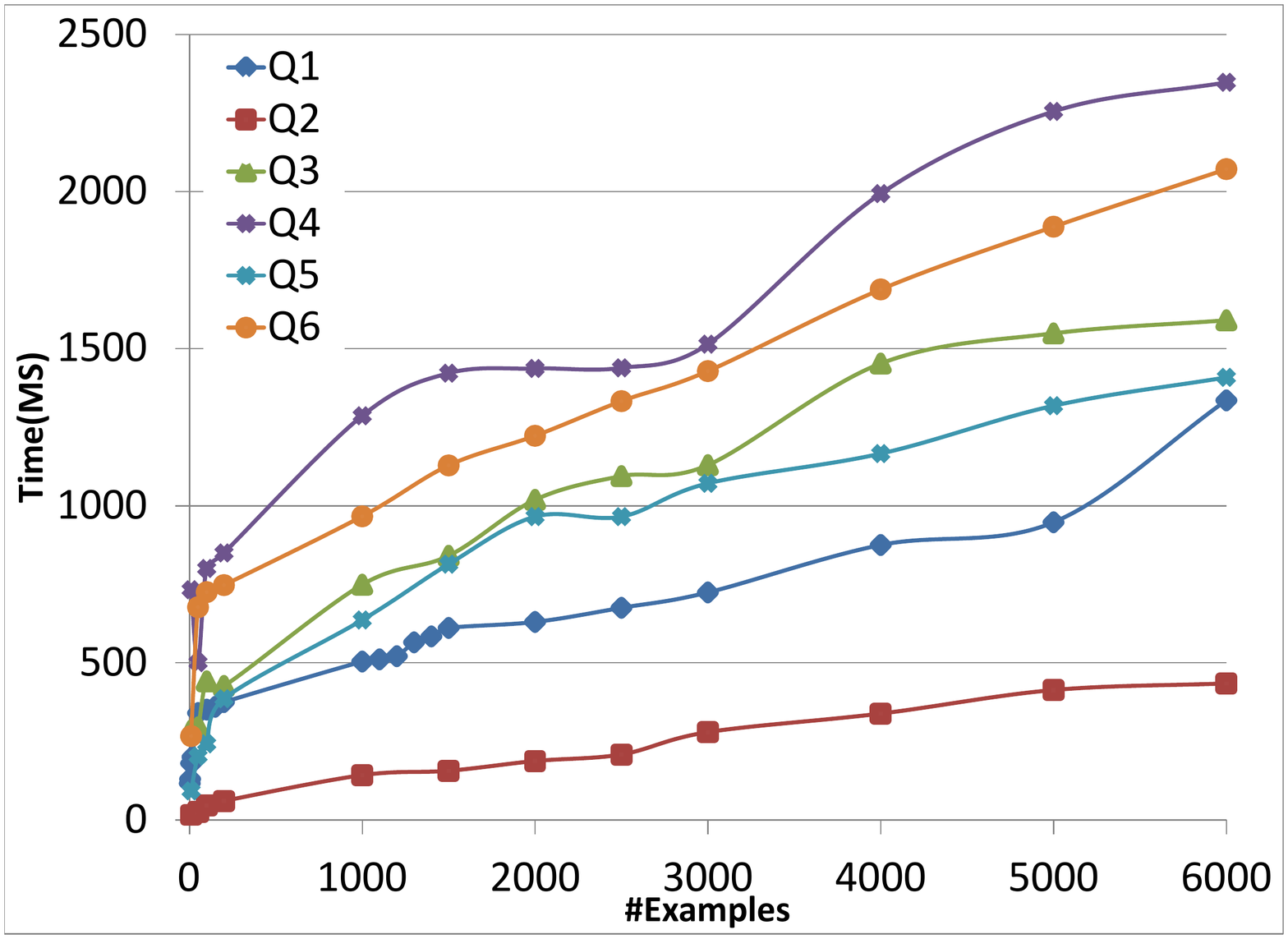}
                 \caption{Queries from \cite{joinQueries}}
                 \label{timesPaperQueries}
         \end{subfigure}%
         \begin{subfigure}[b]{0.26\textwidth}
                 \centering
                 \includegraphics[trim=0cm 2cm 0cm 0cm, width=2.0in]{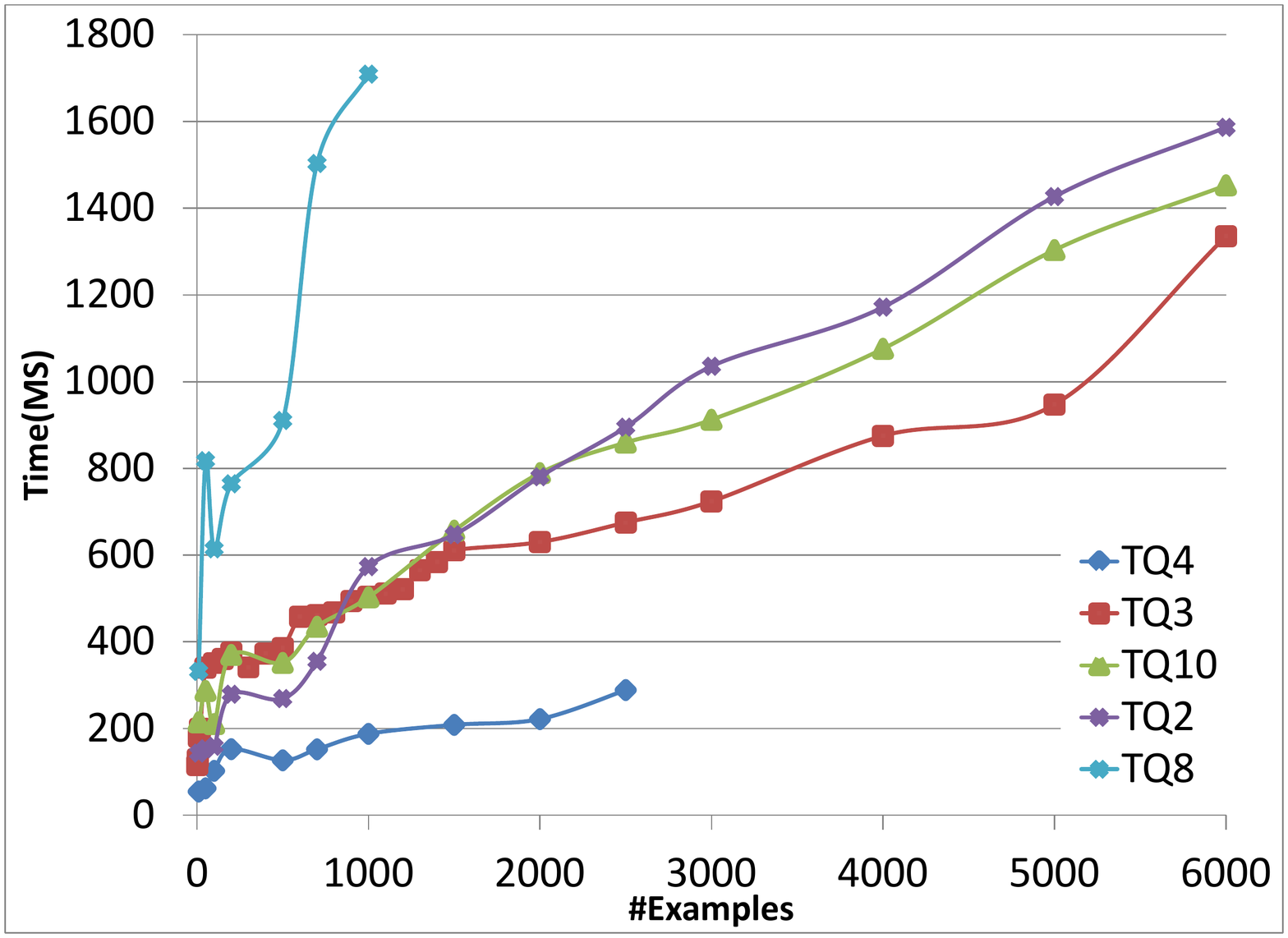}
                 \caption{TPC-H Queries}
                 \label{timesTCP-H}
         \end{subfigure}
\caption{Time of computation in milliseconds as a function of number of examples for $\NX$} \vspace{-4mm}
\label{rowChoosing}
\end{figure}

%\begin{figure}
% %\hspace*{-0.3cm}
% \centering
%                 \centering
%                 \includegraphics[trim=0cm 2cm 0cm 2cm, width=2.7in]{paperQueriesTime.pdf}
%                 \caption{Queries from \cite{joinQueries}}
%                 \label{timesPaperQueries}
%                 \centering
%                 \includegraphics[trim=0cm 2cm 0cm 2cm, width=2.7in]{TCP-H_QueriesTime.pdf}
%                 \caption{TPC-H Queries}
%                 \label{timesTCP-H}
%                 %\vspace{5mm}
%\caption{Time of computation as a function of number of partial matchings for $\NX$} %\vspace{-4mm}
%\label{rowChoosing}
%\end{figure}
%
%
%\begin{figure}
% %\hspace*{-0.3cm}
%  \centering
%                 \includegraphics[trim=0cm 2cm 0cm 2cm, width=2.7in]{paperQueriesTrioTime.pdf}
%                 \caption{Queries from \cite{joinQueries}}
%                 \label{timesPaperQueriesWhy}
%                 \centering
%                 \includegraphics[trim=0cm 2cm 0cm 2cm, width=2.7in]{TCP-H_QueriesTrioTime.pdf}
%                 \caption{TPC-H Queries}
%                 \label{timesTCP-HWhy}
%                 %\vspace{5mm}
%\caption{Time of computation as a function of number of examples for \why} %\vspace{-4mm}
%\label{times}\vspace{-4cm}
%\end{figure}

\begin{figure}
	\hspace*{-0.8cm}
	\centering
	\begin{subfigure}[b]{0.26\textwidth}
		\centering
		\includegraphics[trim=0cm 2cm 0cm 0cm, width=2.0in]{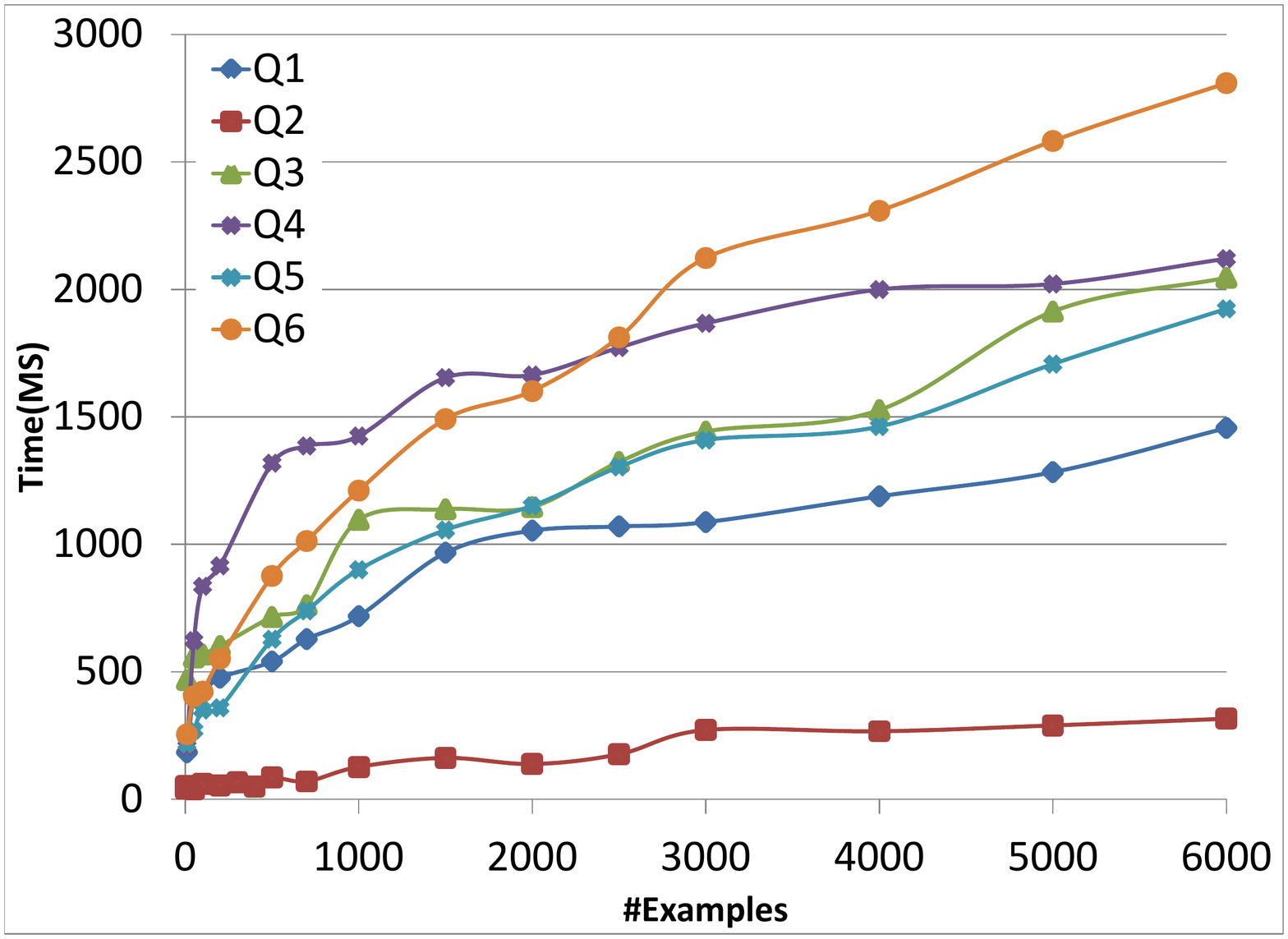}
		\caption{Queries from \cite{joinQueries}}
		\label{timesPaperQueriesWhy}
	\end{subfigure}%
	\begin{subfigure}[b]{0.26\textwidth}
		\centering
		\includegraphics[trim=0cm 2cm 0cm 0cm, width=2.0in]{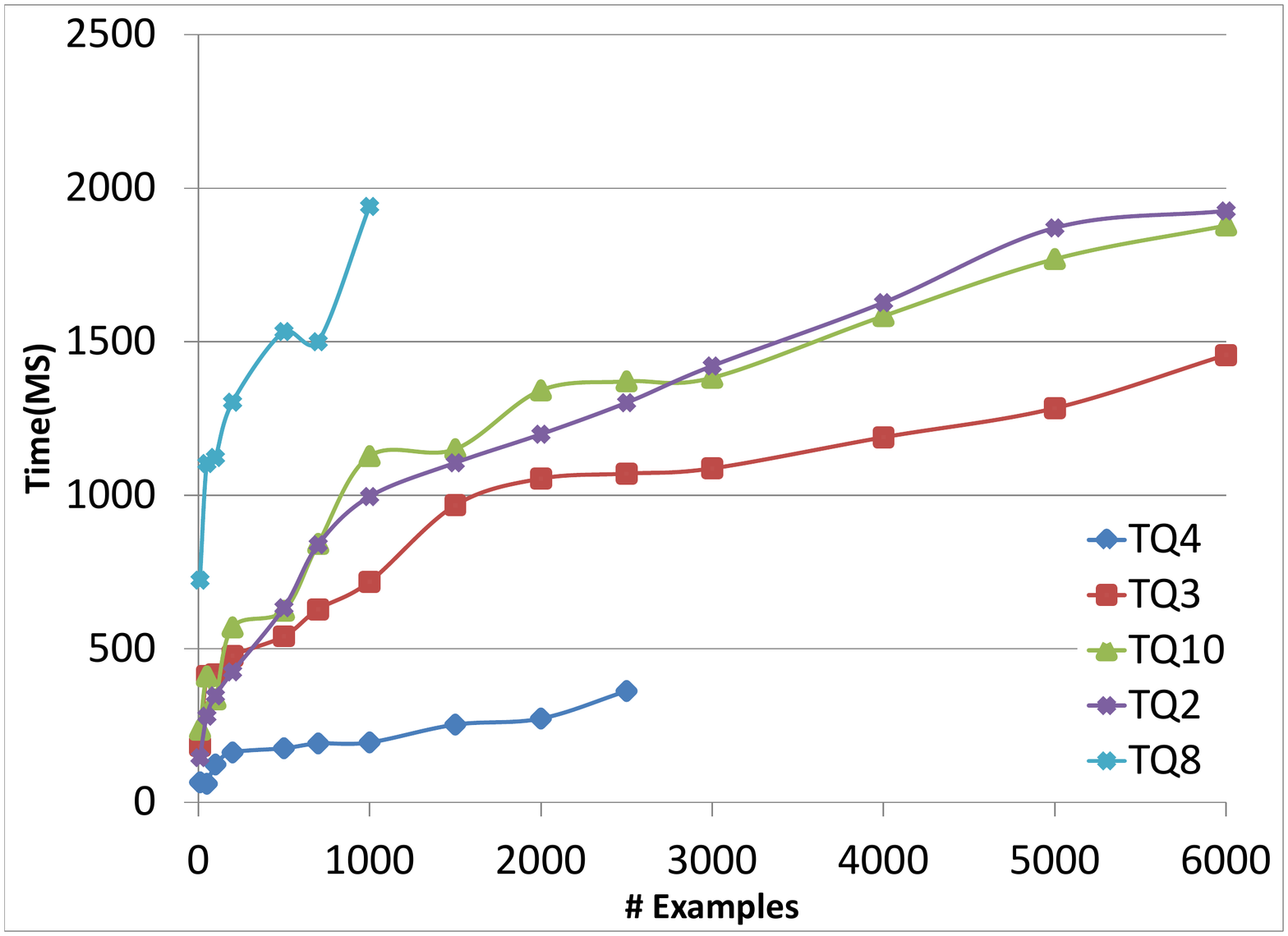}
		\caption{TPC-H Queries}
		\label{timesTCP-HWhy}
	\end{subfigure}
	\caption{Time of computation in milliseconds as a function of number of examples for $Why(X)$} \vspace{-7mm}
\end{figure}

\subsection{Scalability}
Last, we have examined the scalability of our
solution. To this end, we have increased the number of examples up to 6000, which is well beyond a realistic number of
user-provided examples. The results for $\NX$ provenance
and Q1--Q6 are presented in Figure \ref{timesPaperQueries}. The
results exhibit good scalability: the computation time for 6000
examples was 1.3 seconds for Q1 (TQ3), 0.4 seconds for Q2, 2.4
seconds for Q3, 2.3 seconds for Q4 and 1.4 and 2 seconds for Q5 and
Q6 respectively. The performance for the TPC-H queries (Figure
\ref{timesTCP-H}) was similarly scalable: for 6000 examples, the
computation time of TQ2 and TQ10 (which are the queries with the maximum number of head attributes: 8 and 7, resp.) was 1.5 and 1.4 seconds
respectively. The number of examples for queries TQ4, TQ5 and TQ8
was limited due to the queries output size: 2500, 15 and
1000 respectively. The running times for these number of
examples were 0.2, 0.2 and 1.7 seconds respectively.

Next, we have repeated the experiment using $\why$ provenance, and the results appear in Figure 12. In general, the computation time was still fast, and only slightly
slower than the $\NX$ case; this is consistent with our theoretical
complexity analysis. 
%
%The computation times for 6000 tuples were 1.4
%seconds for Q1, 0.3 seconds for Q2, 2 seconds for Q3, 2.1 seconds
%for Q4, 1.9 seconds for Q5 and 2.9 seconds for Q6. The computation for TQ10 and TQ2 has incurred 1.9 seconds;
%for TQ4 with 2500 examples it has incurred 0.35
%seconds. The execution time for TQ8 and TQ5 was approximately the
%same as for the \NX\ case.

%\begin{figure}
%        \centering
%        \includegraphics[trim=0cm 2cm 0cm 0cm, width=2.0in]{matchingQ3.pdf}
%        \caption{Time of computation of Q3 as a function of number of partial matchings for \NX}
%        \label{matchingQ3}
%\end{figure}

\begin{figure}
\vspace{-6mm}
	\centering
	\includegraphics[trim=0cm 2cm 0cm 0cm, width=2in]{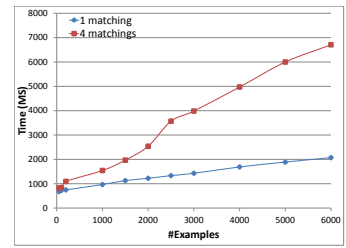}
	\caption{Time of computation in milliseconds of Q6 with a varying number of consistent matchings for \NX}
	\label{matchingQ6}
	\vspace{-5mm}
\end{figure}

\paragraph*{Effect of Tuples Choice}
Recall that Algorithm \ref{algo:Efficient} starts by finding queries that are consistent with two example tuples. We have described a heuristic that chooses the two tuples with
the least number of shared values. The effect of this optimization
is demonstrated in Figure \ref{matchingQ6} for Q6: our choice leads
to a single matching in the graph, as oppose to a random choice of tuples that has led to 4 matchings. The average
overhead of making such a random choice,
instead of using our optimization, was 56\%.
%\daniel{DESCRIBE, HERE OR ELSEWHERE, THE HEURISTIC TO CHOOSE WELL.
%IF DONE ELSEWHERE, REFER TO IT FROM HERE}

%For instance, when the system
%received 1000 examples with the output of Q6, when the system found
%4 matchings the average running time was 1541.66 ms as opposed to
%966.33 ms when only 1 matching was found and 2538.33 ms as opposed
%to 1222.67 ms.

%To conclude this section, our experimental results indicate that in
%most cases, the underlying query is inferred using relatively few
%examples; even if too few examples are provided, the inferred query
%is typically ``not too far" from the underlying query; and the
%execution time of the algorithms is very fast even when presented
%with many examples.

%\input{implementation}

\vspace{-2mm}
\section{Related Work}
\label{sec:related}

There is a large body of literature on learning queries from
examples, in different variants. A first axis of these variants
concerns learning a query whose output {\em precisely} matches the
example (e.g. \cite{qbo,joinQueries}), versus one whose output
contains the example tuples and possibly more (e.g.
\cite{Shen,exampler,Psallidas} and the somewhat different problem in
\cite{zloof}). The first is mostly useful e.g. in a use-case where
an actual query was run and its result, but not the query itself, is
available. This may be the case if e.g. the result was exported and
sent. The second, that we adopt here, is geared towards examples
provided manually by a user, who may not be expected to provide a
full account of the output. Another distinguishing factor between
works in this area is the domain of queries that are inferred; due
to the complexity of the problem, it is typical
(e.g. \cite{joinQueries,Psallidas,Bonifati}) to restrict the attention to
join queries, and many works also impose further restrictions on the
join graph \cite{qbo,DasSarma}. We do not impose such restrictions
and are able to infer complex Conjunctive Queries. Last, there is a
prominent line of work on query-by-example in the context of {\em
data exploration} \cite{Kersten,Bonifati,Abouzied,Bonifati:2014}.
Here users typically provide an initial set of examples, leading to
the generation of a consistent query (or multiple such queries); the
queries and/or their results are presented to users, who may in turn
provide feedback used to refine the queries, and so on. In our
settings, the number of examples required for convergence to the
actual intended query was typically small. In cases where more
examples are needed, an interactive approach is expected to be useful in our setting as well.
%and its study is left for future work.

The fundamental difference between our work and previous work in this area is
the assumed input. Our work is the first, to our knowledge, that
base its inference of queries on explanations that form provenance
information. Leveraging this additional information, we are able to
reach a satisfactory query (1) in a highly complex setting where the
underlying queries includes multiple joins and self-joins, (2) with
no information on the underlying schema beyond relation names and
their number of attributes (in particular no foreign keys are known;
in fact, we do not even need to know the entire input database, but
rather just tuples used in explanations), (3) with only very few
examples (up to 5 were typically sufficient to obtain over 90\%
recall, and less than 20 in all but one case were sufficient to
converge to the actual underlying query), and (4) in split-seconds
for a small number of examples, and in under 3 seconds even with
6000 examples. No previous work, to our knowledge, has exhibited the
combination of these characteristics.

Data Provenance has been extensively studied, for different
formalisms including relational algebra, XML query languages, Nested
Relational Calculus, and functional programs (see e.g.
\cite{trio,GKT-pods07,Userssemiring1,GS13,CheneyProvenance,w3c,ProvenanceBuneman,Olteanu12}).
Many different models have been proposed, and shown useful in a
variety of applications, including program slicing
\cite{slicing}, factorized representation of query results \cite{Olteanu12}, and
``how-to" analysis \cite{Meliou2,Bidoit14query}.%Gatt
We have focused on learning queries from explanations that are either based on the semiring framework or may be expressed using it.
This includes quite a few of the models proposed in the literature, but by no means all of them. Investigating query-by-explanation for
other provenance models is an intriguing direction for future work.

\vspace{-2mm}
\section{Conclusions}
\label{sec:conc}

We have formalized and studied in this paper the problem of
``query-by-explanation", where queries are inferred from example
output tuples and their explanations. We have proposed a generic model, based on the framework of semiring provenance, allowing explanations of varying level of detail and granularity.  We have further presented efficient algorithms that infer conjunctive queries from explanations in one of multiple supported semirings. We have theoretically analysed and experimentally demonstrated the effectiveness of the approach in inferring highly complex queries based on a small number of examples. Intriguing directions for future study include further expressive query languages and additional provenance models including in particular the lineage model. 

%There are many intriguing directions that we intend to pursuit in future work. These include the
%study of query languages beyond Conjunctive Queries, the
%incorporation of user feedback, and the use of explanations based on additional provenance models. In the context of the latter, we in particular intend to further explore the case where explanations are given in Lin(X), as well as cases for which a semiring interpretation may not apply. 

%There are many intriguing directions that we intend to pursuit in future work, including the
%study of query languages beyond Conjunctive Queries, the
%incorporation of user feedback, and the use of explanations based on additional provenance models. 

%In the context of the latter, we in particular intend %to further explore the case where explanations are %given in Lin(X), as well as cases for which a semiring %interpretaion was not proposed. 

%\newpage
% \vspace{-2mm}
%\input{fullMonomial}
% \vspace{-2mm}
%\input{fullPolynomial}
% \vspace{-2mm}
%\input{noCoeffs}
% \vspace{-2mm}
%\input{trio}
% \vspace{-2mm}
%\input{posBool}
% \vspace{-2mm}
%\input{lin}
% \vspace{-2mm}
%\input{conc}

%\paragraph*{Acknowledgments}
%This research was partially supported
%by the Israeli Ministry of Science, by the Isareli Science
%Foundation (ISF), by the Broadcom Foundation and Tel Aviv University
%Authentication Initiative, and by the Advanced ERC grant Modas
%(grant 291071).
%\vspace{-2mm}
%\bibliographystyle{IEEEtran}

%\bibliographystyle{ieeetr}
\bibliographystyle{abbrv}
\small{
\bibliography{bibShort}

\begin{thebibliography}{10}

\bibitem{AHV}
S.~Abiteboul, R.~Hull, and V.~Vianu.
\newblock {\em Foundations of Databases}.
\newblock Addison-Wesley, 1995.

\bibitem{Abouzied}
A.~Abouzied, J.~M. Hellerstein, and A.~Silberschatz.
\newblock Playful query specification with dataplay.
\newblock {\em Proc. VLDB Endow.}, 2012.

\bibitem{slicing}
U.~A. Acar, A.~Ahmed, J.~Cheney, and R.~Perera.
\newblock A core calculus for provenance.
\newblock {\em Journal of Computer Security}, 2013.

\bibitem{Gunopulos1997data}
R.~Agrawal, J.~Gehrke, D.~Gunopulos, and P.~Raghavan.
\newblock Automatic subspace clustering of high dimensional data for data
  mining applications.
\newblock In {\em SIGMOD}, 1998.

\bibitem{Bidoit14query}
N.~Bidoit, M.~Herschel, and K.~Tzompanaki.
\newblock Query-based why-not provenance with nedexplain.
\newblock In {\em EDBT}, 2014.

\bibitem{Bonifati:2014}
A.~Bonifati, R.~Ciucanu, A.~Lemay, and S.~Staworko.
\newblock A paradigm for learning queries on big data.
\newblock Data4U, 2014.

\bibitem{Bonifati}
A.~Bonifati, R.~Ciucanu, and S.~Staworko.
\newblock Interactive join query inference with jim.
\newblock {\em Proc. VLDB Endow.}, 2014.

\bibitem{ProvenanceBuneman}
P.~Buneman, J.~Cheney, and S.~Vansummeren.
\newblock On the expressiveness of implicit provenance in query and update
  languages.
\newblock {\em ACM Trans. Database Syst.}, 2008.

\bibitem{why}
P.~Buneman, S.~Khanna, and W.~Tan.
\newblock Why and where: A characterization of data provenance.
\newblock In {\em ICDT}, 2001.

\bibitem{ChandraMerlin}
A.~K. Chandra and P.~M. Merlin.
\newblock Optimal implementation of conjunctive queries in relational data
  bases.
\newblock In {\em STOC}, 1977.

\bibitem{CheneyProvenance}
J.~Cheney, L.~Chiticariu, and W.~C. Tan.
\newblock Provenance in databases: Why, how, and where.
\newblock {\em Foundations and Trends in Databases}, 2009.

\bibitem{lin}
Y.~Cui, J.~Widom, and J.~L. Wiener.
\newblock Tracing the lineage of view data in a warehousing environment.
\newblock {\em ACM Trans. Database Syst.}, 2000.

\bibitem{DasSarma}
A.~Das~Sarma, A.~Parameswaran, H.~Garcia-Molina, and J.~Widom.
\newblock Synthesizing view definitions from data.
\newblock ICDT, 2010.

\bibitem{icde16}
D.~Deutch and A.~Gilad.
\newblock Qplain: Query by explanation (demo).
\newblock In {\em ICDE}, 2016.
\newblock to appear.

\bibitem{icde15}
D.~Deutch, A.~Gilad, and Y.~Moskovitch.
\newblock selp: Selective tracking and presentation of data provenance.
\newblock In {\em ICDE}, 2015.

\bibitem{Olteanu12}
R.~Fink, L.~Han, and D.~Olteanu.
\newblock Aggregation in probabilistic databases via knowledge compilation.
\newblock {\em PVLDB}, 2012.

\bibitem{Userssemiring1}
F.~Geerts and A.~Poggi.
\newblock On database query languages for k-relations.
\newblock {\em J. Applied Logic}, 2010.

\bibitem{GS13}
B.~Glavic, J.~Siddique, P.~Andritsos, and R.~J. Miller.
\newblock Provenance for data mining.
\newblock In {\em TaPP}, 2013.

\bibitem{Greenicdt09}
T.~J. Green.
\newblock Containment of conjunctive queries on annotated relations.
\newblock In {\em ICDT}, 2009.

\bibitem{GKT-pods07}
T.~J. Green, G.~Karvounarakis, and V.~Tannen.
\newblock Provenance semirings.
\newblock In {\em PODS}, 2007.

\bibitem{hector2002database}
G.-M. Hector, J.~D. Ullman, and J.~Widom.
\newblock {\em Database systems: The complete book}.
\newblock Prentice-Hall, 2002.

\bibitem{Meliou2}
A.~Meliou and D.~Suciu.
\newblock Tiresias: the database oracle for how-to queries.
\newblock In {\em SIGMOD}, 2012.

\bibitem{exampler}
D.~Mottin, M.~Lissandrini, Y.~Velegrakis, and T.~Palpanas.
\newblock Exemplar queries: Give me an example of what you need.
\newblock In {\em VLDB}, 2014.

\bibitem{Psallidas}
F.~Psallidas, B.~Ding, K.~Chakrabarti, and S.~Chaudhuri.
\newblock S4: Top-k spreadsheet-style search for query discovery.
\newblock SIGMOD, 2015.

\bibitem{trio}
A.~D. Sarma, M.~Theobald, and J.~Widom.
\newblock Exploiting lineage for confidence computation in uncertain and
  probabilistic databases.
\newblock In {\em ICDE}, 2008.

\bibitem{Kersten}
T.~Sellam and M.~L. Kersten.
\newblock Meet charles, big data query advisor.
\newblock CIDR, 2013.

\bibitem{Shen}
Y.~Shen, K.~Chakrabarti, S.~Chaudhuri, B.~Ding, and L.~Novik.
\newblock Discovering queries based on example tuples.
\newblock In {\em SIGMOD}, 2014.

\bibitem{tpc}
TPC.
\newblock Tpc benchmarks.
\newblock \url{http://www.tpc.org}.

\bibitem{qbo}
Q.~T. Tran, C.-Y. Chan, and S.~Parthasarathy.
\newblock Query by output.
\newblock In {\em SIGMOD}, 2009.

\bibitem{w3c}
Prov-overview, w3c working group note, 2013.
\newblock http://www.w3.org/TR/prov-overview/.

\bibitem{joinQueries}
M.~Zhang, H.~Elmeleegy, C.~M. Procopiuc, and D.~Srivastava.
\newblock Reverse engineering complex join queries.
\newblock In {\em SIGMOD}, 2014.

\bibitem{zloof}
M.~M. Zloof.
\newblock Query by example.
\newblock In {\em AFIPS NCC}, 1975.

\end{thebibliography}
}
\end{document}